\documentclass{vgtc}                          

\ifpdf
  \pdfoutput=1\relax                   
  \usepackage{graphicx}                
  \DeclareGraphicsExtensions{.pdf,.png,.jpg,.jpeg} 
\else
  \usepackage{graphicx}                
  \DeclareGraphicsExtensions{.eps}     
\fi%

\graphicspath{{figures/}{pictures/}{images/}{./}} 

\usepackage{microtype}                 
\PassOptionsToPackage{warn}{textcomp}  
\usepackage{textcomp}                  
\usepackage{mathptmx}                  
\usepackage{times}                     
\usepackage{cite}                      
\usepackage{tabu}                      
\usepackage{booktabs}                  
\usepackage{amsmath}
\usepackage[svgnames]{xcolor}
\usepackage{algorithm}
\usepackage{algpseudocode}
\usepackage{tabularx}
\usepackage{cuted}

\onlineid{8865}

\vgtccategory{Research}

\vgtcinsertpkg

\title{DeepMetricEye: Metric Depth Estimation in Periocular VR Imagery}

\author{Yitong Sun\thanks{e-mail: yitong.sun@network.rca.ac.uk}\\ %
        \scriptsize Royal College of Art%
\and Zijian Zhou\thanks{e-mail: z.zhou-89@sms.ed.ac.uk}\\ %
     \scriptsize University of Edinburgh%
\and Cyriel Diels\thanks{e-mail: cyriel.diels@rca.ac.uk}\\ %
     \scriptsize Royal College of Art%
\and Ali Asadipour\thanks{e-mail: ali.asadipour@rca.ac.uk}\\ %
     \scriptsize Royal College of Art%
}
\teaser{
  \centering
  \includegraphics[width=\linewidth]{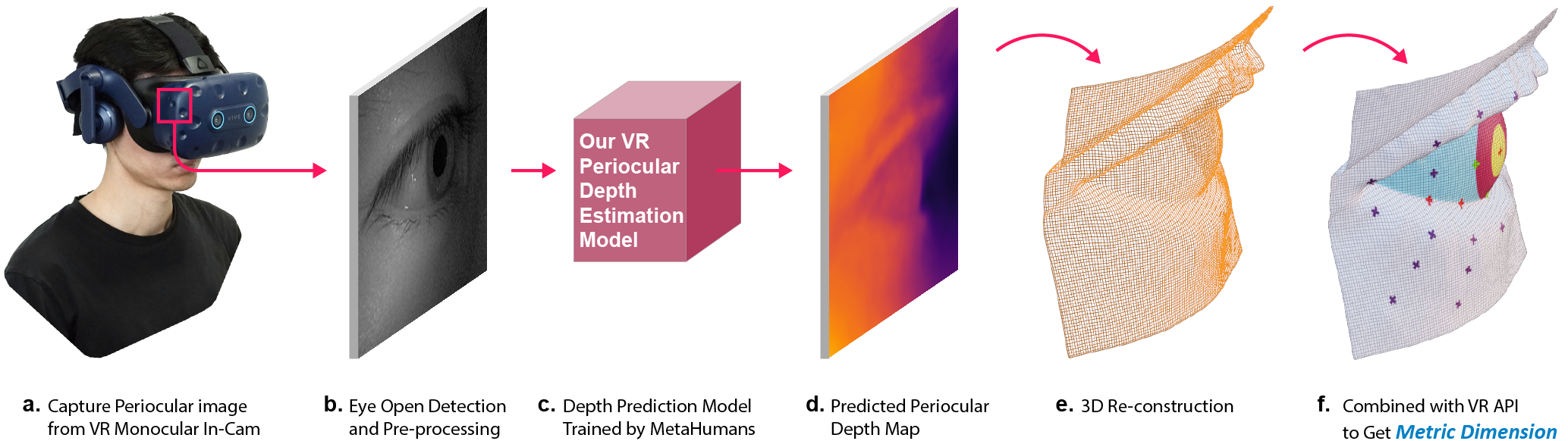}
  \label{fig:teaser}
}

\abstract{Despite the enhanced realism and immersion provided by VR headsets, users frequently encounter adverse effects such as digital eye strain (DES), dry eye, and potential long-term visual impairment due to excessive eye stimulation from VR displays and pressure from the mask. Recent VR headsets are increasingly equipped with eye-oriented monocular cameras to segment ocular feature maps. Yet, to compute the incident light stimulus and observe periocular condition alterations, it is imperative to transform these relative measurements into metric dimensions. To bridge this gap, we propose a lightweight framework derived from the U-Net 3+ deep learning backbone that we re-optimised, to estimate measurable periocular depth maps. Compatible with any VR headset equipped with an eye-oriented monocular camera, our method reconstructs three-dimensional periocular regions, providing a metric basis for related light stimulus calculation protocols and medical guidelines. Navigating the complexities of data collection, we introduce a Dynamic Periocular Data Generation (DPDG) environment based on UE MetaHuman, which synthesises thousands of training images from a small quantity of human facial scan data. Evaluated on a sample of 36 participants, our method exhibited notable efficacy in the periocular global precision evaluation experiment, and the pupil diameter measurement.
}

\CCScatlist{
  \CCScatTwelve{Computing methodologies}{Artificial intelligence}{Computer vision}{Computer vision tasks};
  \CCScatTwelve{Applied computing}{Life and medical sciences}{Consumer health}{}
}

\begin{document}


\firstsection{Introduction}

\maketitle

Virtual Reality (VR) technology has advanced rapidly, offering immersive experiences across applications including gaming, medical, education, and training simulations\cite{dincelli2022immersive, zhang2022artificial}. The recent re-imagination of the 'Digital Universe' concept has illuminated a compelling vision of globally interconnected interactions \cite{mystakidis2022metaverse, ning2023survey}. Despite the continual enhancement in content quality, the use of VR headsets persists in causing physiological discomfort to users, leading to substantially reduced usage duration \cite{keshavarz2022motion, hirzle2022understanding}. A significant proportion of unsatisfactory experiences by VR users can be attributed to digital eye strain (DES), dry eye, and visual impairment resulting from excessive artificial light stimulation from VR displays, as well as periocular swelling, increased intraocular pressure, and muscle displacement induced by the pressure exerted by the headset's face mask \cite{souchet2023narrative}. However, these visual health issues have not yet received attention proportionate to the development of VR technology.

Recent VR headsets, increasingly equipped with eye-oriented monocular cameras, are designed to segment periocular feature maps, annotate the edge of the pupil, and detect gaze direction to enhance content interaction \cite{li2021prospective, fuhl2021teyed}. While these methods offer a preliminary insight into eye activity during VR usage, they are insufficient for establishing connections with medical standards, for instance, the light stimulus calculation protocols and periocular condition medical guidelines, needed for meticulous visual health assessments and advanced user interaction studies \cite{international2018cie, feder2016comprehensive}. The fundamental issue lies in the inability of current methods to convert the segmented 2D relative feature annotations (such as pupil edge segmentation) into spatial metrics (pupil diameter), essential for strict standards. Proposed solutions, for instance, incorporating stereo cameras and depth cameras for metric size acquisition, present substantial challenges in terms of cost, computational power, battery life, and hardware design of VR headsets.

To convert 2D periocular feature annotations into 3D metric dimensions, we propose a framework that only utilises an eye-oriented monocular camera, present in various VR headsets, to estimate the measurable periocular depth map. This framework, built on a U-Net 3+ deep learning backbone, re-optimised by us, aims to accurately estimate depth maps while maintaining the lightweight processing demands suitable for VR deployment \cite{huang2020unet}. To alleviate the difficulty in collecting facial data for training, we introduce a Dynamic Periocular Data Generation (DPDG) environment that leverages a small quantity of real facial scan data to generate thousands of synthetic periocular images and corresponding ground truth depth maps using Unreal Engine (UE) MetaHuman \cite{fang2021metahuman}. \autoref{fig:snapshot} provides the snapshot of this paper.

The main contributions of this study are as follows:

\begin{figure*}[t!]
 \centering
 \includegraphics[width=\textwidth]{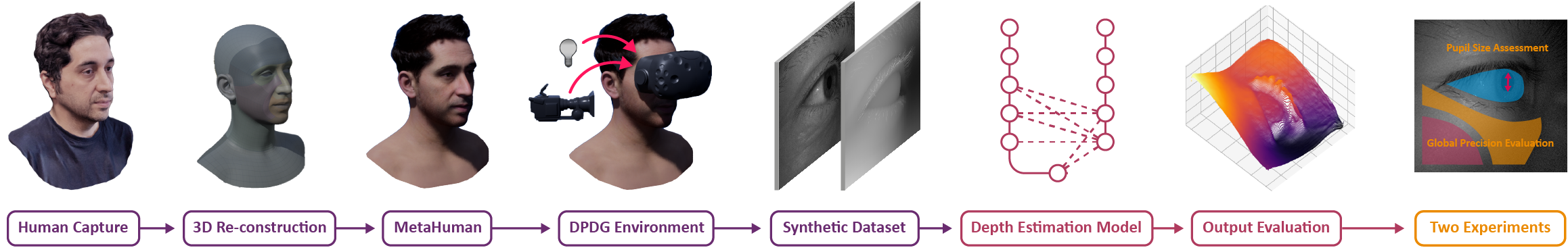}
 \caption{A Snapshot of the paper.}
 \label{fig:snapshot}
\end{figure*}

\begin{enumerate}
    \item  We introduce a lightweight depth estimation framework for VR headsets to reconstruct periocular depth maps. The aim is to provide features' metric size for light stimulus standards calculation and periocular condition monitoring.
    
    \item Addressing the challenge of facial data collection, our DPDG environment, based on UE MetaHuman, generates thousands of periocular training images and depth maps from limited facial scans.
    
    \item We evaluate our method's accuracy and usability with two tasks: 1) evaluating global precision of periocular area , and 2) assessing pupil diameter.

    \item We have open-sourced the DPDG environment, the code and dataset for the depth estimation model, and all metadata from the experiments.
\end{enumerate}

\section{Related Work}
\subsection{Depth Estimation}
Depth estimation is vital in computer vision, with broad applications such as autonomous driving and VR indoor space reconstruction \cite{zhao2020monocular}. Traditional methods rely on stereo vision, estimating depth from disparity across multiple camera views, albeit with high computational demands and need for accurate camera calibration \cite{laga2020survey}. Active depth estimation technologies, using lasers or structured light to generate depth point clouds, offer high accuracy but suffer from slow scanning speeds, high computational needs, and large structure volumes \cite{cheng2018registration, zhang2018high}.

The emergence of image-based depth estimation has sparked research interest \cite{ming2021deep}. Eigen et al. pioneered the deep learning approach for monocular image-based depth estimation using an encoder-decoder network to transform RGB images into depth maps \cite{eigen2014depth}. Subsequent advances include the integration of features like transformers, attention, and residual connections, enhancing model performance but increasing computational demands \cite{li2021revisiting, xu2018structured, chen2019structure, jin2021mono}. 

U-Net, initially designed for medical image segmentation, found application in depth estimation due to its simplicity and practicality \cite{ronneberger2015u, siddique2021u, zhou2020windowed}. U-Net++, proposed by Zongwei et al., introduced dense skip connections but fell short of fully exploiting multiscale information \cite{zhou2018unet++}. To address this, U-Net 3+ was proposed by Huimin et al., using dense skip connections across all encoder and decoder layers, reducing parameter count while improving accuracy \cite{huang2020unet}.

For lightweight, robust depth estimation suitable for VR deployment, we chose U-Net 3+ as our backbone, removing shallow dense skip connections to focus on skin and eye surfaces. Deep decoder layers were weighted differently to control detail decoding, aiming to reduce noise details such as eyebrows and eyelashes.

\subsection{Eye and Periocular Feature Segmentation}
Eye tracking has widespread applications in sectors such as advertising, healthcare, education, and gaming \cite{holmqvist2023eye}. Within VR, gaze detection has been pivotal in identifying users' areas of interest and enhancing content interaction \cite{modi2021review}. Furthermore, it's extensively utilised in research and healthcare for understanding attention, behavioural patterns, and emotional shifts \cite{lisk2020systematic, kang2020identification, lim2020emotion}. Periocular feature segmentation has also been researched to reconstruct facial expressions, acquire iris information, and locate pupils \cite{fuhl2019applicability, wang2017real, fuhl2022pistol}.

Deep learning has substantially improved periocular feature segmentation accuracy and efficiency compared to traditional methods \cite{min2021pupil, chaudhary2019ritnet, yiu2019deepvog}. For instance, the EyeNet framework, with residual units and attention blocks, significantly reduced recognition rate drops caused by flickering and blur noise \cite{kansal2019eyenet}. GazeNet provides end-to-end eye movement detection, focusing on automatic event annotation without pre-labeling \cite{zemblys2019gazenet}.

Despite these advancements, most methods focus on two-dimensional image features, which can't be translated into metrical dimensions in three-dimensional space. This limits the monitoring of eye conditions according to optical standards and medical guidelines, especially given the visual fatigue and impairment caused by VR headsets.

Our method estimates a metrically measurable depth map of the periocular region using monocular images, then combines it with eye segmentation techniques to yield arbitrary feature measurements. This approach allows for effective quantification and monitoring of stimulations on the human eye from VR screen light intrusion and changes caused by the mask pressure on the periocular region.

\subsection{Standards and Guidelines for Periocular Conditions}
Adverse effects on the human eye from VR screens, such as DES, dryness, and visual impairment due to excessive light stimulation, are becoming a focal point of research \cite{souchet2022measuring, hirzle2022understanding}. The face mask's pressure can cause elevated intraocular pressure, nerve pressure, and periorbital swelling \cite{mehrfard2019comparative}. However, despite various studies, no unified method or standard exists for evaluating VR headsets' optical stimuli \cite{hirzle2022understanding, wolffsohn2023tfos, wang2019assessment}. This gap largely stems from the need for accurate pupil area measurements to calculate photoreceptor activation according to International Commission on Illumination (CIE) and American National Standards Institute (ANSI) standards, which current VR headsets' built-in cameras cannot perform \cite{illuminating2014american, international2018cie, international2022cie}. Furthermore, the high contrast in VR headsets complicates stimuli assessment.

Concerning periocular pressure, preliminary studies confirm the negative impacts of varying pressure levels on ocular physiology and immersion \cite{chen2021human}. The comprehensive Adult Medical Eye Evaluation Preferred Practice Pattern (PPP) guidelines outline measurement methods related to periocular symptoms, including muscle displacement and swelling associated with VR usage \cite{feder2016comprehensive}. Yet, the inability of VR's eye-oriented cameras to convert 2D images into the 3D feature information required by PPP poses a significant challenge. 

Our method mitigates this by estimating the periocular area's depth map, converting 2D images into measurable 3D facial reconstructions, and providing spatial metric information aligned with various assessment standards.

\section{Method}
\subsection{Dataset Creation}
\textbf{Objective and Dataset Challenge} \hspace{0.2cm}The aim of our approach is to predict the depth map based on periocular images captured by the monocular camera  within VR headsets, thereby transforming relative facial feature dimensions into metric dimensions. The supervised deep learning training requires two categories of images: 1) periocular images captured by the VR headset's internal camera, and 2) corresponding depth maps used as ground truth for training. However, our analysis of existing periocular image databases revealed a discrepancy as none met the prerequisites for training our model \cite{zanlorensi2022new, zanlorensi2020unconstrained, garbin2019openeds, kim2019nvgaze, vitek2023exploring}. Primarily, databases utilised for gaze, pupil and iris edge detection training do not contain associated depth maps. Moreover, their data were not obtained from the internal camera of VR headsets, which prevents effective calibration of the field of view (FOV), lens distortion, and noise distribution. Additionally, given the compact space within VR headsets, deploying a depth camera to capture depth maps that align with the position of the real camera poses a considerable challenge.

\begin{figure}[!ht]
 \centering
 \includegraphics[width=\columnwidth]{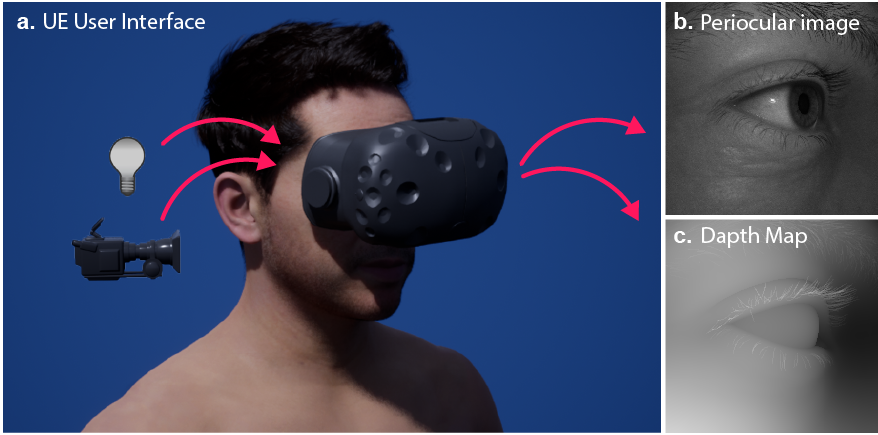}
 \caption{The Dynamic Periocular Data Generation (DPDG) environment. \textbf{a:} Metahumans are worn a specified model of VR headset with faux realistic lighting and camera in UE for periocular image acquisition. \textbf{b:} A captured synthetic periocular image. \textbf{c:} The corresponding depth map.}
 \label{fig:UEUI}
\end{figure}

\begin{figure}[!ht]
 \centering
 \includegraphics[width=\columnwidth]{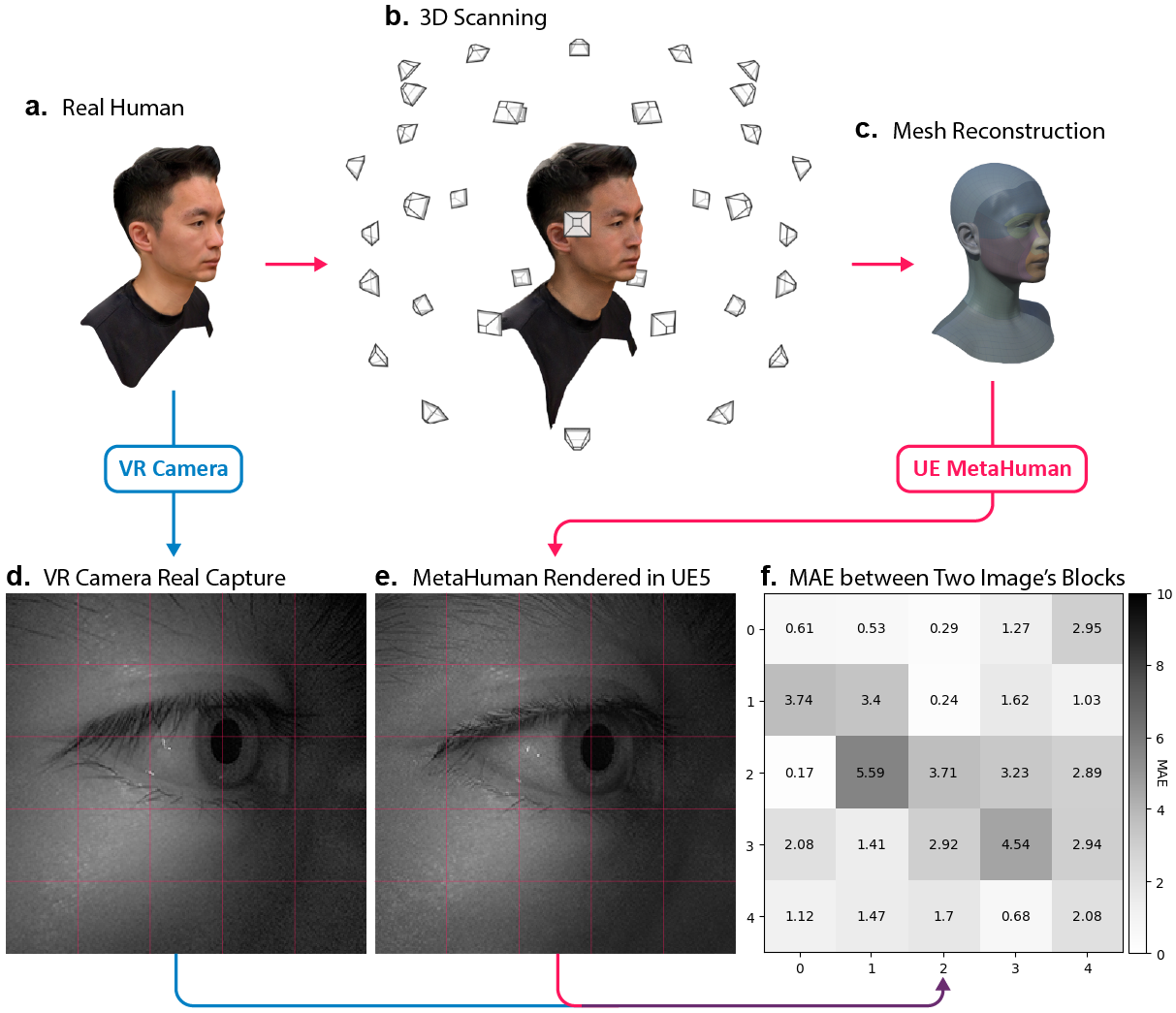}
 \caption{Process of human scan to Metahuman, and synthetic periocular image error quantification and optimisation. \textbf{a:} Real human. \textbf{b:} 3D reconstruction of the face through RealityCapture. \textbf{c:} Converting scanned face mesh to Metahuman. \textbf{d:} Image of the periocular taken with real VR headset. \textbf{e:} Synthetic periocular image through DPDG environment. \textbf{f:} A MAE algorithm iteratively to fine-tune the parameters of the UE cine-camera and UE PointLight to minimise the differences between real and synthetic periocular image.}
 \label{fig:UI}
\end{figure}

\textbf{Dataset Generation} \hspace{0.2cm} In response to these challenges, we created a Dynamic Periocular Data Generation (DPDG) environment using UE's MetaHuman (\autoref{fig:UEUI}). Using this approach, we synthesised virtual humans from real scans, significant mitigating data collection challenges and enhancing dataset diversity. As a state-of-the-art virtual human system, MetaHuman facilitates precise facial feature simulations and can emulate dynamic eye movements, pupil dilation, and blinking. Initially, we procured 68 official MetaHuman avatars and subsequently scanned 52 real humans using RealityCapture to create additional MetaHumans by fitting facial meshes in UE 5 (\autoref{fig:UI} a-c) \cite{Reality_2023}. To enhance the generalisability of the dataset, these 120 MetaHumans underwent a feature-mixing process using MetaHuman Creator, culminating in the generation of 1150 unique avatars \cite{fang2021metahuman}. Using DPDG's feature blending, we combined diverse makeup levels with various skin tones for realism, bolstering dataset robustness. See Appendix B  (FOV2 Row10, FOV4 Row6) and C (Row6) for makeup examples.

\textbf{Demographic Distribution} \hspace{0.2cm} After creating the MetaHumans, it's important to describe the demographic distribution among the original pool of 120, which includes characteristics such as gender, age, skin tone, and ethnicity. The diversified representation within the dataset enhances variety and, consequently, bolsters the model's capability to generalise across different scenarios.

In terms of gender distribution, an equitable balance was sought, resulting in an even split of 50\% for both male and female MetaHumans. Age representation was approached with inclusivity, accounting for individuals in various life stages including youth (15-24 years), young adults (25-44 years), middle-aged individuals (45-64 years), and the elderly (65 years and above). This approach ensures a broad age range within our data.

The variety of skin tones and ethnic backgrounds was another priority, leading to an inclusive set capturing MetaHumans with varied ethnic backgrounds and skin tones. Specifically, the dataset encompasses individuals of Asian, Caucasian, African, Hispanic backgrounds, as well as other ethnicities (mixed race). Skin tones range from fair to dark, with five categories (Fair, Light Medium, Medium, Medium Dark, Dark) considered in our MetaHuman distribution. It should be noted that due to the constraints of the VR headset capturing only in the IR spectrum, considerations for skin tone did not involve hue, but rather the grayscale intensity under the supplementary infrared light-emitting diodes (IR LEDs) illumination.

\autoref{tab:metahuman_distribution} provides a detailed breakdown of the MetaHuman distribution across the aforementioned demographic characteristics:

\begin{table}[h]
\centering
\small
\begin{tabular}{ l l l l }
\hline
\textbf{Attribute} & \textbf{Category} & \textbf{Count} & \textbf{Percentage} \\
\hline
Gender & Male & \textit{60} & \textit{50\%} \\
     & Female & \textit{60} & \textit{50\%} \\
\hline
Age & Youth (15-24 years) & \textit{26} & \textit{21.67\%} \\
     & Young Adult (25-44 years) & \textit{36} & \textit{30\%} \\
     & Middle-aged (45-64 years) & \textit{37} & \textit{30.83\%} \\
     & Elderly (65 years and above) & \textit{21} & \textit{17.5\%} \\
\hline
Skin Tone & Fair & \textit{26} & \textit{21.67\%} \\
     & Light Medium & \textit{32} & \textit{26.67\%} \\
     & Medium & \textit{30} & \textit{25\%} \\
     & Medium Dark & \textit{17} & \textit{14.17\%} \\
     & Dark & \textit{15} & \textit{12.5\%} \\
\hline
Ethnicity & Asian & \textit{24} & \textit{20\%} \\
 & Caucasian & \textit{36} & \textit{30\%} \\
 & African & \textit{24} & \textit{20\%} \\
 & Hispanic & \textit{24} & \textit{20\%} \\
 & Other & \textit{12} & \textit{10\%} \\
\hline
\end{tabular}
\caption{Distribution characteristics of original 120 MetaHumans}
\label{tab:metahuman_distribution}
\end{table}

\begin{figure}[!ht]
 \centering
 \includegraphics[width=\columnwidth]{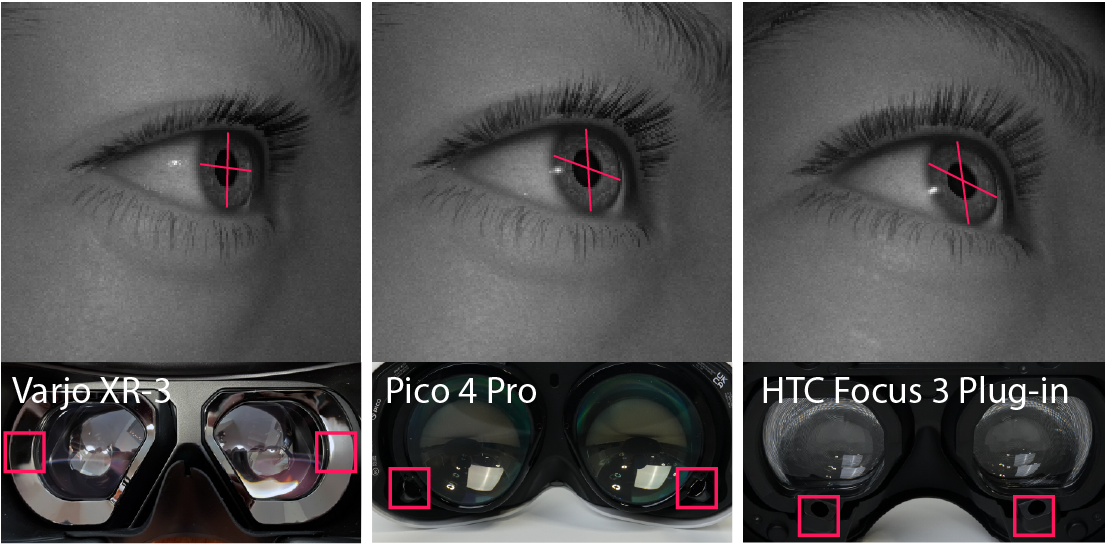}
 \caption{Distribution of monocular camera positions of three VR headsets and corresponding views in DPDG environment. The red cross indicates the perspective reference of the eyeball.}
 \label{fig:headset}
\end{figure}

Utilising UE 5.2, we constructed the DPDG environment. Initially, we secured computer-aided design (CAD) models, camera datasets, and parameters of supplementary IR LED from official sources for various VR headsets, including HTC VIVE Pro Eye, HTC Vive Focus 3, Pico 4 Pro, and Varjo XR-3, to precisely align the spatial position and optical parameters of their internal cameras. \autoref{fig:headset} shows the different distribution of camera positions and corresponding viewpoints in DPDG. Importing CAD models into UE, we placed UE cine-cameras, with parameters adjusted, at the VR internal camera's location. IR LEDs were restored through UE PointLight and were eventually made into Actor blueprints \cite{Engine_2023}. A PostProcessMaterial was applied to the cine-cameras to convert RGB image to depth map image \cite{Unreal_Engine_2023}. To streamline the capture process and boost user-friendliness, we composed an automated collection program based on the blueprint using the UE Editor Utility Widget \cite{Engine_2023_utility}. The program sets the exporting image resolution, selects the VR headset model for collection, sequentially positions the 1150 avatars in the scene, and allows the simulated internal VR camera to individually capture the RGB image and corresponding depth map of each eye. Consequently, 2300 pairs of RGB periocular images and corresponding depth maps per VR headset model were collected.

\textbf{Addressing Robustness} \hspace{0.2cm} Recognising that the unquantified pixel colour disparities between the actual VR capture and the simulation could undermine the robustness of depth map prediction, Despite the highly accurate simulations rendered by MetaHuman, we incorporated a Mean Absolute Error (MAE) metric to quantify differences. This algorithm is used iteratively to fine-tune the parameters of the UE cine-camera and UE PointLight to minimise the differences, thereby optimising the rendering authenticity of the virtual human model.

\textbf{Error Quantification and Optimisation} \hspace{0.2cm} Specifically, we segment each image into $n \times n$ blocks, calculate the average pixel value for each block, and subsequently calculate the MAE between these averages. This block-based approach, rather than calculating the MAE pixel by pixel, mitigates the adverse impact of random noise and minor inconsistencies in facial feature positions on the results, focusing more on the accuracy of the overall colour transition between blocks.

Initially, we divide the actual image $I_r$ and the simulated image $I_s$ into $N$ blocks. For each block $i$, we compute the average pixel values $Avg_{r_i}$ for the real image and $Avg_{s_i}$ for the simulated image as shown in the equation below:

\begin{equation}
Avg_{i} = \frac{1}{|B_i|} \sum_{x \in B_i} I_{i}(x) \quad \text{for} \quad i \in {r, s}
\end{equation}

Here, $B_i$ represents the pixel count in block $i$, while $I_{r_i}(x)$ and $I_{s_i}(x)$ denote the values of pixel $x$ in the real and simulated images in block $i$ respectively.

The MAE for each block, denoted as $MAE_i$, is computed as the absolute difference between the average pixel values of the reference and the simulated image blocks, specifically $Avg_{r_i}$ and $Avg_{s_i}$, respectively. Following this, we aggregate the MAEs for all the blocks to derive the total MAE, represented as $MAE_{total}$. This total MAE is an average of all individual block MAEs and is calculated using the following equation:

\begin{equation}
MAE_{total} = \frac{1}{N} \sum_{i=1}^{N} |Avg_{r_i} - Avg_{s_i}|
\end{equation}

Here, 'N' signifies the total number of blocks under consideration.

Lastly, using a gradient descent approach, we adjust the parameters of the UE cine-camera's random noise ($\theta_{\text {noise}}$) and the UE PointLight's intensity ($\theta_{\text {light}}$) to minimise the overall MAE. In each iteration, we update these parameters as a vector $\theta = [\theta_{\text {noise}}, \theta_{\text {light}}]^T$ based on the gradient of the loss function $L(\theta)$, which corresponds to $MAE_{total}$, as expressed in the equation below:

\begin{equation}
\begin{array}{l}
\left[\begin{array}{l}
\theta_{\text {noise}} \\
\theta_{\text {light}}
\end{array}\right]=\left[\begin{array}{l}
\theta_{\text {noise}} \\
\theta_{\text {light}}
\end{array}\right]-\alpha \nabla L\left(\left[\begin{array}{l}
\theta_{\text {noise}} \\
\theta_{\text {light}}
\end{array}\right]\right)
\end{array}
\end{equation}

Here, $\alpha$ is the learning rate, governing the magnitude of parameter updates at each step. We continue with this update procedure until $L(\theta)$ reaches a minimum value that meets our threshold (0.5\%), or the predetermined number of iterations is reached (is set to 100 steps).

\textbf{Error Analysis} \hspace{0.2cm} To affirm our algorithm's efficacy, we undertook iterative optimisation using four different VR headsets encompassed within our dataset. A representation of the terminal optimisation utilising the Varjo headset is depicted in \autoref{fig:UI} d-f. Our algorithm consistently reduced disparities between simulated and actual images, keeping the MAE below an acceptable threshold (average within 1\% of the 0-255 pixel colour spectrum), affirming its effectiveness.

Given the depth map range spans from 20mm to 90mm, a 1\% error corresponds to an average deviation of 0.7mm in depth estimation under optimal conditions. This degree of error is deemed acceptable for numerous measurement applications within VR. Task-specific errors will be addressed within the \nameref{sec:experimrnt} section.

\textbf{Limitations} \hspace{0.2cm} It is important to recognise that the current iteration of MetaHuman is incapable of accurately reconstructing eyebrow and eyelash features, which accounts for why the MAE values of certain blocks are comparatively high. Nonetheless, the overall performance of our optimisation approach offers a robust basis for periocular depth map estimation.

\begin{figure*}[t!]
 \centering
 \includegraphics[width=\textwidth]{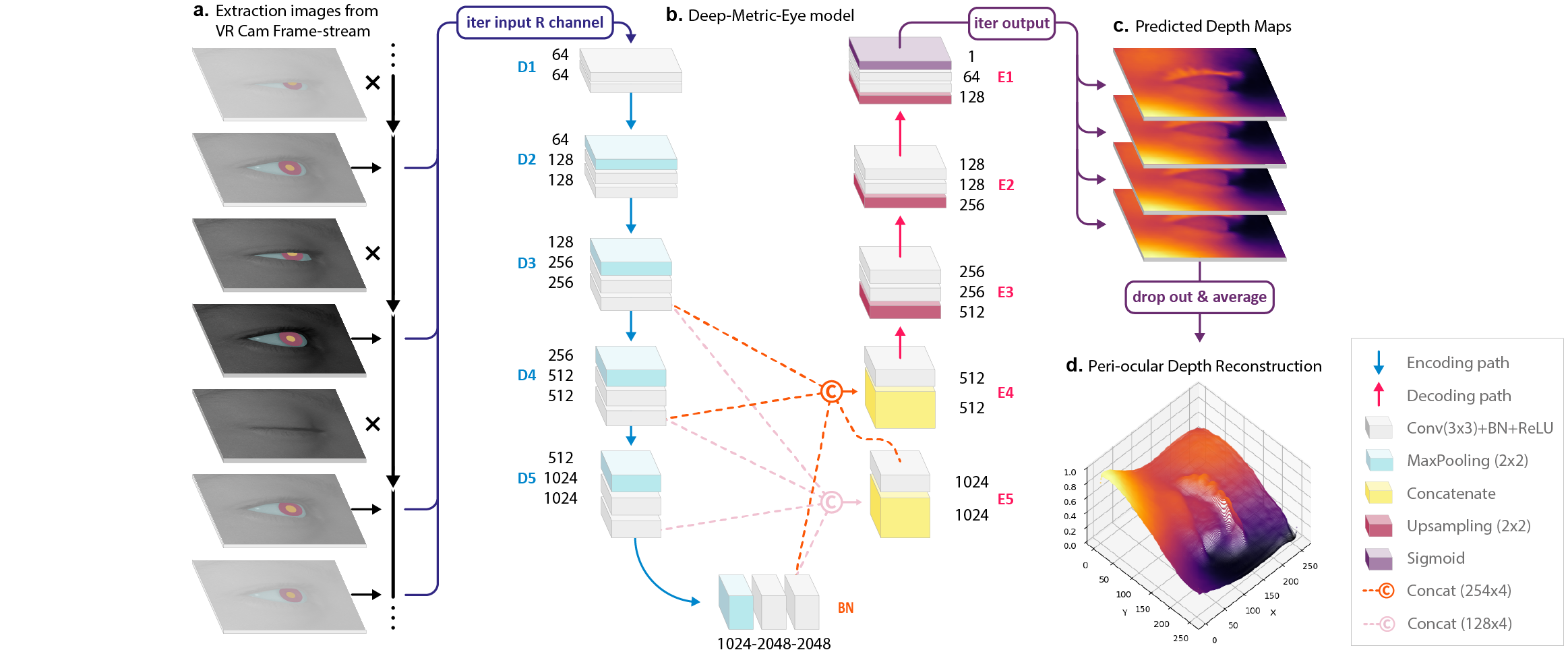}
 \caption{Flowchart of proposed depth estimation framework. \textbf{a:} Initial phase involves detection of open-eye state and gaze direction using VR headset's API, from which a sequence of periocular images consistent with open-eye position and gaze direction is extracted from the video stream. \textbf{b:} The red channel of extracted images are iteratively input to the depth estimation model, an lightweighted and optimised U-Net 3+ variant with a 5-layer symmetrical encoder-decoder structure. The model omits shallow dense skip connections to diminish the negative impact of intricate details, such as pupils, eyelashes, and eyebrow regions, on the smooth transitions of the depth map, thereby prioritising deep semantics. The numbers indicate the depth dimensions of the tensors. \textbf{c:} The output depth maps undergo a two-standard-deviation outlier elimination and pixel averaging to produce \textbf{d}, the final periocular depth estimation.}
 \label{fig:process}
\end{figure*}

\subsection{Model Development}
\textbf{Model Overview} \hspace{0.2cm} A three-stage periocular depth estimation model is proposed in this study. \autoref{fig:process} illustrates the framework. In the initial stage, we utilise the VR headset's eye edge detection API to extract a series of images from the video frame flow, ensuring they align with specific eye-opening conditions. Subsequently, the selected images are iteratively input into a deep learning network, which is composed of a 5-layer symmetrical encoder-decoder, for the generation of corresponding predicted depth maps. Finally, these depth maps undergo outlier exclusion via a two-standard-deviation threshold and averaging to produce the ultimate periocular depth prediction. Further application of this model, in conjunction with the eye feature recognition API of the VR headset, allows for the calculation of metric dimensions of arbitrary features.

\subsubsection{Data Pre-processing}
\textbf{Multi-image Collection Strategy} \hspace{0.2cm} Given the nature of wearing a VR headset, the relative position of the eye and the camera remains constant. In periocular depth estimation tasks, the process is executed as a long interval operation where accuracy of the single operation is given priority over high frame-rate processing. Therefore, we design a multi-image collection strategy for the data preparation phase. This strategy utilises built-in eye feature segmentation APIs of various VR systems to judge and collect multiple images consistent in eyelid opening degree and gaze direction. Once multiple estimated depth maps are obtained, outlier detection and processing are performed using the Median Absolute Deviation (MAD) method. MAD, a robust outlier detection method based on median absolute deviation, is less prone to extreme value impact than the standard Z-score \cite{curtis2016mystery}. Utilising this method effectively eliminates outliers in depth maps, and averages are taken for each pixel to improve prediction robustness.

\textbf{VR API Integration} \hspace{0.2cm} All current VR headsets with eye-oriented cameras feature eye segmentation APIs for pupil and gaze detection. Given user privacy concerns, these APIs are proprietary. We obtained testing permissions through NDAs with HTC and PICO. For Varjo XR-3 PC VR, direct access to eye images captured by the camera is available. To ensure the complete reproducibility of the entire process, the open-source human eye segmentation framework, EllSeg, proposed by Kothari et al., was also employed as a direct-use implementation for Varjo XR-3 \cite{kothari2021ellseg}.

\textbf{Image Capture Condition} \hspace{0.2cm} The extent of eyelid opening is crucial for tasks like facial expression recognition, fatigue detection, and depth estimation. Our method for determining eye openness is a cost-efficient computational strategy, which doesn't necessitate a specific detection orientation and has been robustly tested within systems by Meta and HTC. The approach begins by segmenting and curve-fitting the upper and lower eyelid contours. Subsequently, it identifies the midpoint of both curves, evaluating the amplitude intervals and extremities of their movement to deduce the level of eye openness. In our study, images are captured when the gaze direction is straight ahead and the eyelids are naturally at their widest extent.

The procedure during the data preparation phase is detailed in the pseudocode provided in Appendix A, as Algorithm \ref{alg:pre-precessing}.

\textbf{Model Consideration} \hspace{0.2cm} Transitioning from data preparation, our key challenge was designing a deep learning model suitable for the VR context, especially for standalone headsets with their computational and memory constraints. We considered architectures like DenseNet, ResNet, MonoDepth, Transformer, and Deeper Depth Prediction, all of which are renowned for depth estimation tasks\cite{iandola2014densenet, he2016deep, godard2017unsupervised, vaswani2017attention, laina2016deeper}. While these models have demonstrated significant performance in their respective domains, their inherent complexity and high parameter count make them less suitable for the constrained resources of VR applications. For instance, DenseNet's dense connections between all layers and ResNet's multiple stacked residual blocks, although beneficial for their specific use cases, escalate the model complexity and parameter count, posing challenges in the VR context. Increased parameters entail augmented memory and computational demands, a challenge for standalone VR headsets bereft of potent GPUs. Moreover, the inherent architectural complexity augments energy consumption, potentially hastening battery depletion and inducing latency disruptions in real-time interactions.

\begin{figure*}[t!]
 \centering
 \includegraphics[width=\textwidth]{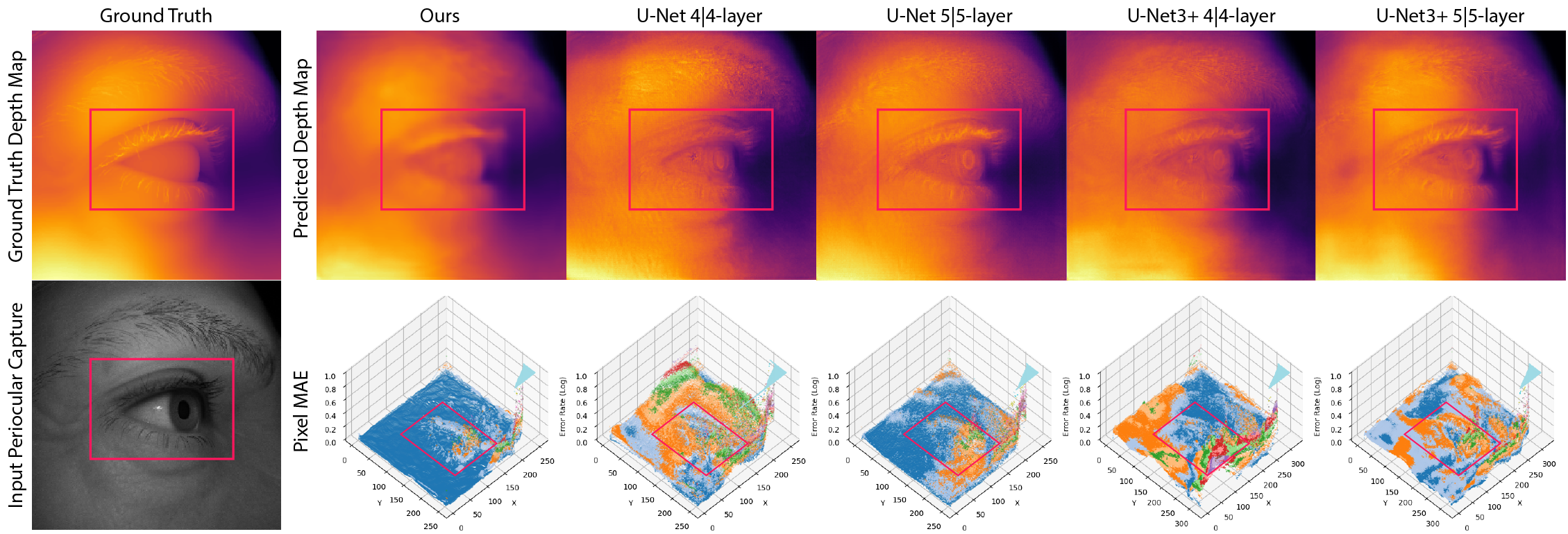}
 \caption{A three-dimensional visualisation of the MAE for the depth estimation from our model and four original structured U-Net models. The MAE values have been log-transformed to emphasise the differences. The x and y-axes represent the length and width of the image, while the z-axis represents the log-transformed MAE values. The visualisation demonstrates our model's superior depth estimation performance, particularly in intricate areas such as around the pupils and cheekbones. It highlights the benefits of our model's focus on deep semantics, and the shortcomings of models giving equal attention to both surface details and deeper semantics.}
 \label{fig:compare}
\end{figure*}

\textbf{Re-imagined U-Net Architecture for VR} \hspace{0.2cm}  In the evolution from PC VR to standalone VR systems, there's a pronounced need for a lightweight yet robust deep learning model. The original U-Net architecture, commendable in its own right, faced challenges. Specifically, the classic 4-layer encoder-decoder network struggled to adequately abstract conflicting colour features, such as hair and pupils, as can be observed in the details of Figure \ref{fig:compare}. To address this, we reconfigured the U-Net blueprint. We judiciously omitted shallow skip connections, which, although beneficial for certain applications, introduced potential noise due to their emphasis on low-level features. Instead, we favoured a system that separately handles abstraction and concrete expressiveness. Further, in the deeper layers of the network, weights were assigned based on the depth of the network to each encoding layer participating in skip connections, controlling the transmission of feature details and abstraction. By augmenting the classic 4-layer structure with an additional layer, transforming it into a 5-layer symmetric structure, our model bolstered its capacity to accurately capture the nuanced depth variations inherent to the eye region. While PC-based depth estimation solutions often veer towards increased complexity, our re-imagined U-Net strikes a balance between efficiency and performance, offering a promising approach for similar challenges, especially in the context of resource-constrained VR devices.

\subsubsection{Model Training}
\textbf{Training Dataset Preparation} \hspace{0.2cm} Model training was performed separately for each VR headset model. For each model, 2300 pairs of periocular RGB images and corresponding ground truth depth maps, both of 256x256 resolution, were randomly apportioned into training, validation, and testing subsets, adhering to a 70\%-15\%-15\% distribution. As the camera within the VR is equipped with a filter film that matches the wavelength of the supplementary IR LEDs to avoid screen light interference, it results in its effective light-sensitive channel being the red channel only. Therefore, during the data pre-processing stage, in order to be consistent with the VR camera and to further reduce the number of model parameters, the RGB images used for training were channel split, with only the red single channel data being extracted for training and the green and blue channels being discarded.

\textbf{Model Architecture Detail} \hspace{0.2cm} As shown in \autoref{fig:process} b, our model comprises a five-layer down-sampling convolutional encoder, a bottleneck layer, and a five-layer up-sampling convolutional decoder. Each layer of the five-layer encoder and the bottleneck layer employs a convolution operation with a 3x3 kernel, stride of 2, and padding of 1, followed by a ReLU activation function and batch normalisation. This sequence is repeated once within each layer. Pooling operations with a pool size and stride of 2 succeed each encoder layer, further reducing the spatial dimensions.

In the decoder part, we introduce skip connections to enable feature-sharing from the third, fourth, fifth layer of encoder and bottleneck layer with the corresponding fourth and fifth layer in the decoder. Additionally, to regulate the propagation of features during decoding, we assign varying weights to the encoder layers based on their depth. This strategic weighting ensures that our model captures and propagates the necessary level of detail during the encoding-decoding process. The specific weights applied are represented by the following function:

\begin{algorithm}
\caption{Weights Set for Model Skip Concatenations}
\label{alg:concatWeights}
\begin{algorithmic}[1]
\Function{skipConcatenate}{$E3, E4, E5, BN$}
    \State $D5 \gets concat(0.1E3, 0.8E4, 1.0E5, 1.0BN)$
    \State $D4 \gets concat(0.2E3, 0.5E4, 0.8BN, 1.0D5)$
    \State \textbf{return} $D4, D5$
\EndFunction
\end{algorithmic}
\end{algorithm}

The third to fifth decoder layers initially undergo a bilinear up-sampling operation with a scale of 2, succeeded by two sets of 3x3 convolutions (stride 2, padding 1), ReLU activation, and batch normalisation. Finally, we apply a 1x1 convolution, followed by a Sigmoid activation function. All convolutional layers adopt He initialisation for weights and zero initialisation for biases.

\textbf{Loss Function} \hspace{0.2cm} The chosen loss function for our depth estimation task is the Reverse Huber Loss (BerHu). The BerHu loss function, given by Equation (\ref{eq:berHu}), exhibits different behaviours contingent on the magnitude of the error, \(x\), which corresponds to the disparity between the predicted and actual depth values. The parameter \(c\) denotes a threshold, which has been set at 24\% of the maximum absolute error within a mini batch after our iterative testing.

\begin{equation}
\label{eq:berHu}
  L_{\text{berHu}}(x) = 
  \begin{cases} 
    |x| & \text{if } |x| \leq c, \\
    \frac{x^2 + c^2}{2c} & \text{otherwise},
  \end{cases}
\end{equation}

For errors that do not exceed \(c\), BerHu aligns with the L1 loss function, thereby functioning as an absolute error. This linear component renders the loss function resilient to outliers, mitigating their impact on the model's learning trajectory and thereby enhancing model robustness.

In the scenario where errors exceed \(c\), BerHu mirrors the L2 loss function, acting as a squared error. This quadratic aspect imposes more substantial penalties on larger errors. The consequential drive to minimise these errors results in the model paying enhanced attention to such instances.

The distinct dual character of the BerHu loss function enables effective management of both minor and major prediction errors, demonstrating its suitability for depth estimation tasks.

\textbf{Regularisation} \hspace{0.2cm} We adopt several strategies to counteract over-fitting. Dropout layers, with a dropout rate of 0.5, are inserted into the fifth layer of the encoder and the bottleneck layer. Additionally, we introduce an early stopping mechanism that halts training if there is no improvement in the validation loss after 20 consecutive epochs. Training is set for a maximum of 150 epochs, starting with an initial learning rate of 1e-4.

\begin{table*}[!t]
\centering
\begin{tabular}{c c c c c c c c c c }
\hline
  &  & & Depth Error($\downarrow$) & & & Depth Accuracy($\uparrow$) & & Model Size($\downarrow$) \\
\hline
Model & AbsRel & SqRel & RMSE & RMSElog & $\delta<1.25$ & $\delta<1.25^2$ & $\delta<1.25^3$ & Parameters\\
\hline
Ours-VIVE Pro Eye & 0.039 & 0.012 & 0.013 & 1.009 & 0.979 & 0.985 & 0.986 & 28.8M\\
Ours-VIVE Focus3 & 0.041 & 0.014 & 0.021 & 0.921 & 0.971 & 0.981 & 0.983 & 28.8M\\
Ours-Pico 4 Pro & 0.037 & 0.012 & 0.018 & 1.109 & 0.969 & 0.986 & 0.982 & 28.8M\\
Ours-Varjo XR-3 & 0.035 & 0.011 & 0.014 & 1.122 & 0.976 & 0.984 & 0.987 & 28.8M\\
\hline
Ours Avg. & \textbf{0.038} & \textbf{0.012} & \textbf{0.017} & \textbf{1.040} & \textbf{0.974} & \textbf{0.984} & \textbf{0.985} & 28.8M\\
\hline
U-Net 4|4-layer\cite{ronneberger2015u} & 0.648 & 0.161 & 0.182 & 1.830 & 0.244 & 0.558 & 0.818 & 31.1M\\
U-Net 5|5-layer & 0.064 & 0.021 & 0.022 & 1.155 & 0.963 & 0.982 & 0.978 & 53.6M\\
U-Net3+ 4|4-layer\cite{zhou2020windowed} & 0.121 & 0.029 & 0.043 & 1.119 & 0.957 & 0.980 & 0.984 & \textbf{26.9M}\\
U-Net3+ 5|5-layer & 0.092 & 0.018 & 0.038 & 1.023 & 0.965 & 0.983 & 0.979 & 44.6M\\
\hline
\end{tabular}
\caption{Comparative analysis of our model and the original structured U-Net models in terms of depth error, depth accuracy, and parameters. The depth error and parameters are preferred to be lower (represented by $\downarrow$), and depth accuracy is preferred to be higher (represented by $\uparrow$). Results are given for different VR headset datasets and the average performance is also provided. Our model consistently outperforms the U-Net models, showing superior performance across different headset datasets and maintaining a comparable model size to the 4-layer U-Net3+ model.}
\label{tab:model_comparison}
\end{table*}

\subsubsection{Model Evaluation}

\textbf{Evaluation and Comparative Analysis} \hspace{0.2cm} We synthesised perspective-appropriate training datasets for various VR headsets in the DPDG environment and subjected our model to training and evaluation. To present an integrated perspective on the model's performance, we averaged the model performance metrics across different headsets and compared them with four original structured U-Net models as the benchmark. These encompass the U-Net and U-Net 3+ models with both 4-layer and 5-layer symmetrical structures.

Multiple metrics were employed to measure depth estimation error (lower is better) and accuracy (higher is better), including Absolute Relative Difference (AbsRel), Squared Relative Difference (SqRel), Root Mean Square Error (RMSE), Logarithmic Root Mean Square Error (RMSElog), and the accuracy under thresholds of $1.25$, $1.25^2$, and $1.25^3$.

The results, summarised in Table \ref{tab:model_comparison}, clearly demonstrate that, compared to the four original U-Net models, our model consistently outperforms in terms of both depth estimation error and accuracy across all tested VR headsets, indicating substantial generalisability. Moreover, it is worth noting that our model achieves a deeper layer structure with a parameter count comparable to that of the 4-layer symmetrical U-Net 3+ model, proving to be more efficient than the other three models.

\textbf{Visualisation of Errors} For a detailed performance insight, we visualised depth estimation results from a representative image in the validation set (\autoref{fig:compare}). This involved comparing our model's estimations with the ground truth and calculating the MAE for each normalised depth value. The log-transformed MAE values, shown in a 3D plot, indicate our model's superior accuracy, especially around the pupils and cheekbones, suggesting its utility for tasks like measuring pupil diameter. In comparison, other models struggle with regions like the pupil due to their equal focus on surface and deeper semantics.

\textbf{Analysis Summary} \hspace{0.2cm} On the whole, our model manifests a robust performance. In comparison with the original U-Net models, our model demonstrates superior depth estimation accuracy and lower error across multiple VR headsets. Moreover, the visualisation further emphasises the advantages of our model, especially in estimating the depth of complex regions of detail and transition such as eyelashes, eyebrows, and image boundaries.

\section{Experiment}
\label{sec:experimrnt}
\begin{figure}[!ht]
 \centering
 \includegraphics[width=\columnwidth]{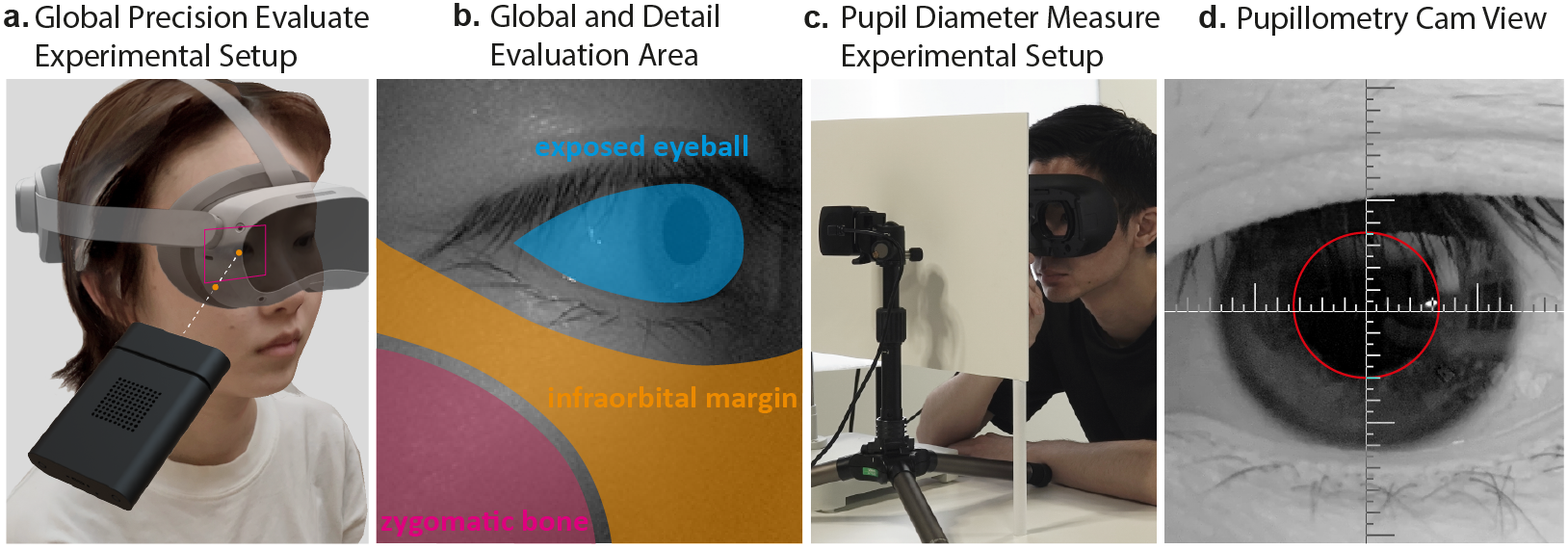}
 \caption{Experimental Setup\textbf{a:} ORBBEC Femto ToF camera and PICO 4 Pro VR headset alignment for periocular region imaging. \textbf{b:} Illustration of the regions used for global and regional accuracy evaluation. \textbf{c:} Setup for the pupil measurement experiment. \textbf{d:} Pupillometry Camera view for pupil diameter measurement.}
 \label{fig:expt}
\end{figure}

To demonstrate our VR depth estimation model's practical applicability, we conducted two experiments. In the \textbf{Global Precision Evaluation}, we employed the ORBBEC Femto Time of Flight (ToF) camera, chosen for its renowned near-limit accuracy in commercial devices, boasting a depth map accuracy error of only 0.2\%. By comparing the depth data from participants' periocular regions captured by this ToF camera with our model's estimations, this experiment aimed to evaluate our model's global region error, assessing its potential as a feasible alternative to high-precision equipment. The \textbf{Pupil Diameter Measurement} experiment evaluated the model's capability to measure fine details. Under constant lighting, we measured participants' pupil sizes and contrasted them with sizes from our model's depth maps. Additionally, prediction time for each depth map was recorded. \autoref{fig:expt} depicts the experimental setup.

36 participants (18-60 years, balanced by gender) with a variety of skin tones participated in our study. They were health-screened, with no vision impairments. To ensure uniformity, glasses and contact lenses were removed, and of these participants, 12 wore makeup. All participants gave informed consent after understanding the study's procedures and potential discomforts. They were informed of their right to withdraw anytime. The study was approved by the Institutional Review Board of Royal College of Art, adhering to ethical standards.

\subsection{Global Precision Evaluation}
In the first experiment, each participant underwent two periocular data acquisitions for both the left and right eyes. Initially, participants were equipped with a PICO 4 Pro VR headset. We deployed our proposed model within the VR system using the \textit{Pytorch Android Mobile} deployment process to estimate depth maps at a 256x256 pixel resolution. To establish a consistent baseline and counteract camera distortion, we transformed the depth map to absolute spatial coordinates using the equation:

\begin{equation}\label{eq:5}
(X, Y, Z) = ((x - c_x) * s * D_p / f_x, (y - c_y) * s * D_p / f_y, s * D_p)
\end{equation}

where $D_p$ represents the pixel value in the depth map, while $s$ is a conversion scaling factor used to convert pixel values into the real-world depth, Z. The coordinates $x$ and $y$ represent the pixel location in the image, and $(X, Y, Z)$ are the corresponding coordinates in three-dimensional space. $c_x$ and $c_y$ are the positions of the optical centre of the camera lens in the image coordinate system, typically located at the centre of the image. The terms $f_x$ and $f_y$ denote the focal lengths of the camera.

For the second acquisition, the Femto ToF camera was aligned to mirror the viewpoint of the PICO headset's internal camera and was positioned 20cm away from the participant's eye (its minimum effective focal distance), capturing periocular depth data at a 256x256 pixel resolution. The absolute coordinates were derived from the official camera API.

Upon concluding the measurements, we amassed 72 valid samples. The pixel-based Mean Absolute Error (\(MAE_p\)) between the two depth maps was calculated to be 1.68mm, with a standard deviation (\(sd\)) of 1.22mm across all pixels. The predominant errors, mainly observed in the eyebrow and eyelash regions, arise from the Femto camera's limitations in effectively capturing hair. Additionally, our model's strategic abstraction of these regions to achieve surface smoothness might also be a contributing factor to the discrepancies.

However, when focusing on regions of paramount importance for medical standards, such as the exposed eyeball, infraorbital margin, and zygomatic bone, our model showcased commendable precision. The \(MAE_p\) values for these regions were 0.63mm, 0.74mm, and 0.57mm, respectively, with \(sd\) values of 0.27mm, 0.35mm, and 0.24mm. These findings underline the potential of our approach in practical, medically-relevant scenarios. The detailed comparison results and regional discrepancies for a part of the samples are presented in Appendix D. Notably, despite the makeup variations across participants, the results remained consistent, affirming the robustness of our model, which was trained on datasets inclusive of diverse makeup levels. On average, the model required 8.11 seconds to predict a depth map.

\subsection{Pupil Diameter Measurement}

For the secondary experiment, participants were positioned in a controlled environment with consistent illumination, oriented towards a uniformly frosted cardboard sheet. Centrally, an aperture was incorporated into the cardboard to facilitate a calibrated \textit{Baumer VCXU.2-123C} pupillometry camera, purposed for frontally gauging participants' pupil diameters. Concurrently, participants were equipped solely with the HTC Focus 3 eye tracker. The main body of the VR headset, tethered via a USB type-c cable, was strategically placed laterally. Depth map prediction was executed via the \textit{Pytorch Android Mobile} deployment within the VR system. The pupillometry camera and HTC eye tracker accessory operated in tandem, producing reference diameter measurements and predictive depth maps. Following this, the absolute pupil diameter was ascertained employing the VR system's intrinsic pupil segmentation API in conjunction with equation \ref{eq:5}. The resulting discrepancy was quantified against the reference value by computing the absolute differential between the ground truth pupil size and the size deduced from the depth map.

Notably, while corneal refraction possesses the potential to perturb pupil diameter measurements, preliminary simulations, constructed via Zemax modelling of an idealised human ocular structure, suggest that refractive aberrations owing to corneal interference are inconsequential under our experimental conditions, especially when assessing pupil diameter at oblique angles. Comprehensive specifications and resultant data from these simulations are elaborated upon in Appendix E.

Subsequent to experimental completion, a corpus of 72 valid measurement samples was curated, delineating pupil diameters spanning from 3.61mm to 4.48mm, averaging at 5.17mm. The divergence between authentic and estimated pupil diameters yielded an average (\(\mu\)) of 0.33mm, accompanied by a standard deviation (\(sd\)) of 0.14mm. The ensuing percentage discrepancy totaled 6.38\%. The mean computational time required for depth map prediction was registered at 7.31 seconds.

\subsection{Analysis of Results}
The experimental outcomes validate the efficacy and precision of our periocular depth estimation framework in both feature measurement and micro-scale feature assessment. In the Global Precision Evaluation experiment, the model showcased high accuracy in regions critical for health standards calculation, despite diminished precision in areas covered by hair. However, in the Pupil Diameter Measurement experiment, a slight increase in error was observed. We hypothesise this discrepancy results from the limited pixel area occupied by the pupil in low-resolution images, potentially leading to imprecise pupil edge segmentation due to pixel aliasing. Addressing this limitation by enhancing acquisition resolution or employing curve fitting could refine the accuracy of micro-scale feature estimations. These findings underscore the framework's viability for advanced eye state monitoring in VR settings, aligning with light stimulation standards and medical guidelines.

\section{Conclusion}
In summary, our study presents a composite framework for periocular measurable depth estimation capable of efficiently and accurately predicting spatial metric dimensions for eye region features using an eye-oriented monocular VR camera. To mitigate the challenges of facial data collection for training the model, we introduce our DPDG environment, which can generate synthetic periocular datasets for various VR headsets using MetaHuman. Through having conducted two practical experiments, the Global Precision Evaluation and the Pupil Diameter Measurement, our model has proven its robust capabilities in spatial depth estimation and detailed small-scale feature assessment. Our aim is to bridge the gap between eye and periocular state changes during VR immersion and light stimulation standards and medical guidelines, thereby facilitating deep and comprehensive monitoring, with the ultimate goal of effectively quantifying stimuli to mitigate the harm inflicted on human eyes by current VR headsets usage.

\textbf{Limitation and Future Work} \hspace{0.2cm} Despite the successful deployment of our framework in VR, the volume of parameters remains substantial. We hypothesise that this could be due to the profusion of superfluous details in the eyelash and eyebrow areas, which may have significantly consumed network fitting efficiency and drastically increased the parameter count. To address this, a potentially effective solution might involve deleting eyelashes and eyebrows when creating ground truth depth maps within the DPDG, while preserving the details in the RGB images for model training. Another limitation is the exclusive focus on participants not wearing glasses. The presence of glasses could pose heightened challenges in depth estimation due to reflections, lens distortions, and obstructions, warranting exploration in future studies. Although the current straight-forward gaze acquisition suffices for most scenarios, we envision potential dynamic acquisition in the future by integrating 3D ocular surface reconstruction\cite{yiu2019deepvog, dierkes2018novel}. Lastly, While pixel-based edge segmentation offers speed advantages, it can result in jagged edges when measuring small-scale features such as the pupil, leading to reduced accuracy. Transitioning to curve fitting might provide a more refined approach, enhancing precision.

\section{Open Science}
To promote further research into eye health within VR, we have made our DPDG dataset synthesis environment and the depth estimation model / code available to the public through the provided GitHub link\footnote{https://github.com/sunyitong/DPDG-Env}. For the experimental metadata containing facial information, please contact the authors and sign an NDA before accessing, to ensure the facial privacy rights of the participants.


\bibliographystyle{unsrt}   

{\small
\bibliography{template}       

\begin{thebibliography}{10}

\bibitem{dincelli2022immersive}
Ersin Dincelli and Alper Yayla.
\newblock Immersive virtual reality in the age of the metaverse: A
  hybrid-narrative review based on the technology affordance perspective.
\newblock {\em The journal of strategic information systems}, 31(2):101717,
  2022.

\bibitem{zhang2022artificial}
Zixuan Zhang, Feng Wen, Zhongda Sun, Xinge Guo, Tianyiyi He, and Chengkuo Lee.
\newblock Artificial intelligence-enabled sensing technologies in the
  5g/internet of things era: from virtual reality/augmented reality to the
  digital twin.
\newblock {\em Advanced Intelligent Systems}, 4(7):2100228, 2022.

\bibitem{mystakidis2022metaverse}
Stylianos Mystakidis.
\newblock Metaverse.
\newblock {\em Encyclopedia}, 2(1):486--497, 2022.

\bibitem{ning2023survey}
Huansheng Ning, Hang Wang, Yujia Lin, Wenxi Wang, Sahraoui Dhelim, Fadi Farha,
  Jianguo Ding, and Mahmoud Daneshmand.
\newblock A survey on the metaverse: The state-of-the-art, technologies,
  applications, and challenges.
\newblock {\em IEEE Internet of Things Journal}, 2023.

\bibitem{keshavarz2022motion}
Behrang Keshavarz and John~F Golding.
\newblock Motion sickness: current concepts and management.
\newblock {\em Current opinion in neurology}, 35(1):107--112, 2022.

\bibitem{hirzle2022understanding}
Teresa Hirzle, Fabian Fischbach, Julian Karlbauer, Pascal Jansen, Jan
  Gugenheimer, Enrico Rukzio, and Andreas Bulling.
\newblock Understanding, addressing, and analysing digital eye strain in
  virtual reality head-mounted displays.
\newblock {\em ACM Transactions on Computer-Human Interaction (TOCHI)},
  29(4):1--80, 2022.

\bibitem{souchet2023narrative}
Alexis~D Souchet, Domitile Lourdeaux, Alain Pagani, and Lisa Rebenitsch.
\newblock A narrative review of immersive virtual reality’s ergonomics and
  risks at the workplace: cybersickness, visual fatigue, muscular fatigue,
  acute stress, and mental overload.
\newblock {\em Virtual Reality}, 27(1):19--50, 2023.

\bibitem{li2021prospective}
Fan Li, Ching-Hung Lee, Shanshan Feng, Amy Trappey, and Fazal Gilani.
\newblock Prospective on eye-tracking-based studies in immersive virtual
  reality.
\newblock In {\em 2021 IEEE 24th International Conference on Computer Supported
  Cooperative Work in Design (CSCWD)}, pages 861--866. IEEE, 2021.

\bibitem{fuhl2021teyed}
Wolfgang Fuhl, Gjergji Kasneci, and Enkelejda Kasneci.
\newblock Teyed: Over 20 million real-world eye images with pupil, eyelid, and
  iris 2d and 3d segmentations, 2d and 3d landmarks, 3d eyeball, gaze vector,
  and eye movement types.
\newblock In {\em 2021 IEEE International Symposium on Mixed and Augmented
  Reality (ISMAR)}, pages 367--375. IEEE, 2021.

\bibitem{international2018cie}
International~Commission on~Illumination.
\newblock Cie system for metrology of optical radiation for iprgc-influenced
  responses to light, 2018.

\bibitem{feder2016comprehensive}
Robert~S Feder, Timothy~W Olsen, Bruce~E Prum, C~Gail Summers, Randall~J Olson,
  Ruth~D Williams, and David~C Musch.
\newblock Comprehensive adult medical eye evaluation preferred practice
  pattern{\textregistered} guidelines.
\newblock {\em Ophthalmology}, 123(1):P209--P236, 2016.

\bibitem{huang2020unet}
Huimin Huang, Lanfen Lin, Ruofeng Tong, Hongjie Hu, Qiaowei Zhang, Yutaro
  Iwamoto, Xianhua Han, Yen-Wei Chen, and Jian Wu.
\newblock Unet 3+: A full-scale connected unet for medical image segmentation.
\newblock In {\em ICASSP 2020-2020 IEEE International Conference on Acoustics,
  Speech and Signal Processing (ICASSP)}, pages 1055--1059. IEEE, 2020.

\bibitem{fang2021metahuman}
Zhixin Fang, Libai Cai, and Gang Wang.
\newblock Metahuman creator the starting point of the metaverse.
\newblock In {\em 2021 International Symposium on Computer Technology and
  Information Science (ISCTIS)}, pages 154--157. IEEE, 2021.

\bibitem{zhao2020monocular}
Chaoqiang Zhao, Qiyu Sun, Chongzhen Zhang, Yang Tang, and Feng Qian.
\newblock Monocular depth estimation based on deep learning: An overview.
\newblock {\em Science China Technological Sciences}, 63(9):1612--1627, 2020.

\bibitem{laga2020survey}
Hamid Laga, Laurent~Valentin Jospin, Farid Boussaid, and Mohammed Bennamoun.
\newblock A survey on deep learning techniques for stereo-based depth
  estimation.
\newblock {\em IEEE transactions on pattern analysis and machine intelligence},
  44(4):1738--1764, 2020.

\bibitem{cheng2018registration}
Liang Cheng, Song Chen, Xiaoqiang Liu, Hao Xu, Yang Wu, Manchun Li, and Yanming
  Chen.
\newblock Registration of laser scanning point clouds: A review.
\newblock {\em Sensors}, 18(5):1641, 2018.

\bibitem{zhang2018high}
Song Zhang.
\newblock High-speed 3d shape measurement with structured light methods: A
  review.
\newblock {\em Optics and lasers in engineering}, 106:119--131, 2018.

\bibitem{ming2021deep}
Yue Ming, Xuyang Meng, Chunxiao Fan, and Hui Yu.
\newblock Deep learning for monocular depth estimation: A review.
\newblock {\em Neurocomputing}, 438:14--33, 2021.

\bibitem{eigen2014depth}
David Eigen, Christian Puhrsch, and Rob Fergus.
\newblock Depth map prediction from a single image using a multi-scale deep
  network.
\newblock {\em Advances in neural information processing systems}, 27, 2014.

\bibitem{li2021revisiting}
Zhaoshuo Li, Xingtong Liu, Nathan Drenkow, Andy Ding, Francis~X Creighton,
  Russell~H Taylor, and Mathias Unberath.
\newblock Revisiting stereo depth estimation from a sequence-to-sequence
  perspective with transformers.
\newblock In {\em Proceedings of the IEEE/CVF International Conference on
  Computer Vision}, pages 6197--6206, 2021.

\bibitem{xu2018structured}
Dan Xu, Wei Wang, Hao Tang, Hong Liu, Nicu Sebe, and Elisa Ricci.
\newblock Structured attention guided convolutional neural fields for monocular
  depth estimation.
\newblock In {\em Proceedings of the IEEE conference on computer vision and
  pattern recognition}, pages 3917--3925, 2018.

\bibitem{chen2019structure}
Xiaotian Chen, Xuejin Chen, and Zheng-Jun Zha.
\newblock Structure-aware residual pyramid network for monocular depth
  estimation.
\newblock {\em arXiv preprint arXiv:1907.06023}, 2019.

\bibitem{jin2021mono}
Yifan Jin, Lei Yu, Zhong Chen, and Shumin Fei.
\newblock A mono slam method based on depth estimation by densenet-cnn.
\newblock {\em IEEE Sensors Journal}, 22(3):2447--2455, 2021.

\bibitem{ronneberger2015u}
Olaf Ronneberger, Philipp Fischer, and Thomas Brox.
\newblock U-net: Convolutional networks for biomedical image segmentation.
\newblock In {\em Medical Image Computing and Computer-Assisted
  Intervention--MICCAI 2015: 18th International Conference, Munich, Germany,
  October 5-9, 2015, Proceedings, Part III 18}, pages 234--241. Springer, 2015.

\bibitem{siddique2021u}
Nahian Siddique, Sidike Paheding, Colin~P Elkin, and Vijay Devabhaktuni.
\newblock U-net and its variants for medical image segmentation: A review of
  theory and applications.
\newblock {\em Ieee Access}, 9:82031--82057, 2021.

\bibitem{zhou2020windowed}
Lipu Zhou and Michael Kaess.
\newblock Windowed bundle adjustment framework for unsupervised learning of
  monocular depth estimation with u-net extension and clip loss.
\newblock {\em IEEE Robotics and Automation Letters}, 5(2):3283--3290, 2020.

\bibitem{zhou2018unet++}
Zongwei Zhou, Md~Mahfuzur Rahman~Siddiquee, Nima Tajbakhsh, and Jianming Liang.
\newblock Unet++: A nested u-net architecture for medical image segmentation.
\newblock In {\em Deep Learning in Medical Image Analysis and Multimodal
  Learning for Clinical Decision Support: 4th International Workshop, DLMIA
  2018, and 8th International Workshop, ML-CDS 2018, Held in Conjunction with
  MICCAI 2018, Granada, Spain, September 20, 2018, Proceedings 4}, pages 3--11.
  Springer, 2018.

\bibitem{holmqvist2023eye}
Kenneth Holmqvist, Saga~Lee {\"O}rbom, Ignace~TC Hooge, Diederick~C Niehorster,
  Robert~G Alexander, Richard Andersson, Jeroen~S Benjamins, Pieter Blignaut,
  Anne-Marie Brouwer, Lewis~L Chuang, et~al.
\newblock Eye tracking: empirical foundations for a minimal reporting
  guideline.
\newblock {\em Behavior research methods}, 55(1):364--416, 2023.

\bibitem{modi2021review}
Nandini Modi and Jaiteg Singh.
\newblock A review of various state of art eye gaze estimation techniques.
\newblock {\em Advances in Computational Intelligence and Communication
  Technology: Proceedings of CICT 2019}, pages 501--510, 2021.

\bibitem{lisk2020systematic}
Stephen Lisk, Ayesha Vaswani, Marian Linetzky, Yair Bar-Haim, and Jennifer~YF
  Lau.
\newblock Systematic review and meta-analysis: Eye-tracking of attention to
  threat in child and adolescent anxiety.
\newblock {\em Journal of the American Academy of Child \& Adolescent
  Psychiatry}, 59(1):88--99, 2020.

\bibitem{kang2020identification}
Jiannan Kang, Xiaoya Han, Jiajia Song, Zikang Niu, and Xiaoli Li.
\newblock The identification of children with autism spectrum disorder by svm
  approach on eeg and eye-tracking data.
\newblock {\em Computers in biology and medicine}, 120:103722, 2020.

\bibitem{lim2020emotion}
Jia~Zheng Lim, James Mountstephens, and Jason Teo.
\newblock Emotion recognition using eye-tracking: taxonomy, review and current
  challenges.
\newblock {\em Sensors}, 20(8):2384, 2020.

\bibitem{fuhl2019applicability}
Wolfgang Fuhl, David Geisler, Wolfgang Rosenstiel, and Enkelejda Kasneci.
\newblock The applicability of cycle gans for pupil and eyelid segmentation,
  data generation and image refinement.
\newblock In {\em Proceedings of the IEEE/CVF International Conference on
  Computer Vision Workshops}, pages 0--0, 2019.

\bibitem{wang2017real}
Kang Wang and Qiang Ji.
\newblock Real time eye gaze tracking with 3d deformable eye-face model.
\newblock In {\em Proceedings of the IEEE International Conference on Computer
  Vision}, pages 1003--1011, 2017.

\bibitem{fuhl2022pistol}
Wolfgang Fuhl, Daniel Weber, and Shahram Eivazi.
\newblock Pistol: Pupil invisible supportive tool to extract pupil, iris, eye
  opening, eye movements, pupil and iris gaze vector, and 2d as well as 3d
  gaze.
\newblock {\em arXiv preprint arXiv:2201.06799}, 2022.

\bibitem{min2021pupil}
Nasro Min-Allah, Farmanullah Jan, and Saleh Alrashed.
\newblock Pupil detection schemes in human eye: a review.
\newblock {\em Multimedia Systems}, 27(4):753--777, 2021.

\bibitem{chaudhary2019ritnet}
Aayush~K Chaudhary, Rakshit Kothari, Manoj Acharya, Shusil Dangi, Nitinraj
  Nair, Reynold Bailey, Christopher Kanan, Gabriel Diaz, and Jeff~B Pelz.
\newblock Ritnet: Real-time semantic segmentation of the eye for gaze tracking.
\newblock In {\em 2019 IEEE/CVF International Conference on Computer Vision
  Workshop (ICCVW)}, pages 3698--3702. IEEE, 2019.

\bibitem{yiu2019deepvog}
Yuk-Hoi Yiu, Moustafa Aboulatta, Theresa Raiser, Leoni Ophey, Virginia~L
  Flanagin, Peter Zu~Eulenburg, and Seyed-Ahmad Ahmadi.
\newblock Deepvog: Open-source pupil segmentation and gaze estimation in
  neuroscience using deep learning.
\newblock {\em Journal of neuroscience methods}, 324:108307, 2019.

\bibitem{kansal2019eyenet}
Priya Kansal and Sabarinathan Devanathan.
\newblock Eyenet: Attention based convolutional encoder-decoder network for eye
  region segmentation.
\newblock In {\em 2019 IEEE/CVF International Conference on Computer Vision
  Workshop (ICCVW)}, pages 3688--3693. IEEE, 2019.

\bibitem{zemblys2019gazenet}
Raimondas Zemblys, Diederick~C Niehorster, and Kenneth Holmqvist.
\newblock gazenet: End-to-end eye-movement event detection with deep neural
  networks.
\newblock {\em Behavior research methods}, 51:840--864, 2019.

\bibitem{souchet2022measuring}
Alexis~D Souchet, St{\'e}phanie Philippe, Domitile Lourdeaux, and Laure Leroy.
\newblock Measuring visual fatigue and cognitive load via eye tracking while
  learning with virtual reality head-mounted displays: A review.
\newblock {\em International Journal of Human--Computer Interaction},
  38(9):801--824, 2022.

\bibitem{mehrfard2019comparative}
Arian Mehrfard, Javad Fotouhi, Giacomo Taylor, Tess Forster, Nassir Navab, and
  Bernhard Fuerst.
\newblock A comparative analysis of virtual reality head-mounted display
  systems.
\newblock {\em arXiv preprint arXiv:1912.02913}, 2019.

\bibitem{wolffsohn2023tfos}
James~S Wolffsohn, Gareth Lingham, Laura~E Downie, Byki Huntjens, Takenori
  Inomata, Saleel Jivraj, Emmanuel Kobia-Acquah, Alex Muntz, Karim
  Mohamed-Noriega, Sotiris Plainis, et~al.
\newblock Tfos lifestyle: impact of the digital environment on the ocular
  surface.
\newblock {\em The ocular surface}, 28:213--252, 2023.

\bibitem{wang2019assessment}
Yan Wang, Guangtao Zhai, Sichao Chen, Xiongkuo Min, Zhongpai Gao, and Xuefei
  Song.
\newblock Assessment of eye fatigue caused by head-mounted displays using
  eye-tracking.
\newblock {\em Biomedical engineering online}, 18:1--19, 2019.

\bibitem{illuminating2014american}
Illuminating Engineering Society~ANSI/IES RP-3-13.
\newblock American national standard practice on lighting for educational
  facilities, 2014.

\bibitem{international2022cie}
International~Commission on~Illumination.
\newblock Light and lighting — integrating lighting — non-visual effects,
  2022.

\bibitem{chen2021human}
Yumiao Chen, Xin Wang, and Huijia Xu.
\newblock Human factors/ergonomics evaluation for virtual reality headsets: a
  review.
\newblock {\em CCF Transactions on Pervasive Computing and Interaction},
  3(2):99--111, 2021.

\bibitem{zanlorensi2022new}
Luiz~A Zanlorensi, Rayson Laroca, Diego~R Lucio, Lucas~R Santos, Alceu~S
  Britto~Jr, and David Menotti.
\newblock A new periocular dataset collected by mobile devices in unconstrained
  scenarios.
\newblock {\em Scientific Reports}, 12(1):17989, 2022.

\bibitem{zanlorensi2020unconstrained}
Luiz~A Zanlorensi, Hugo Proen{\c{c}}a, and David Menotti.
\newblock Unconstrained periocular recognition: Using generative deep learning
  frameworks for attribute normalization.
\newblock In {\em 2020 IEEE International Conference on Image Processing
  (ICIP)}, pages 1361--1365. IEEE, 2020.

\bibitem{garbin2019openeds}
Stephan~J Garbin, Yiru Shen, Immo Schuetz, Robert Cavin, Gregory Hughes, and
  Sachin~S Talathi.
\newblock Openeds: Open eye dataset.
\newblock {\em arXiv preprint arXiv:1905.03702}, 2019.

\bibitem{kim2019nvgaze}
Joohwan Kim, Michael Stengel, Alexander Majercik, Shalini De~Mello, David Dunn,
  Samuli Laine, Morgan McGuire, and David Luebke.
\newblock Nvgaze: An anatomically-informed dataset for low-latency, near-eye
  gaze estimation.
\newblock In {\em Proceedings of the 2019 CHI conference on human factors in
  computing systems}, pages 1--12, 2019.

\bibitem{vitek2023exploring}
Matej Vitek, Abhijit Das, Diego~Rafael Lucio, Luiz~Antonio Zanlorensi, David
  Menotti, Jalil~Nourmohammadi Khiarak, Mohsen~Akbari Shahpar, Meysam
  Asgari-Chenaghlu, Farhang Jaryani, Juan~E. Tapia, Andres Valenzuela, Caiyong
  Wang, Yunlong Wang, Zhaofeng He, Zhenan Sun, Fadi Boutros, Naser Damer,
  Jonas~Henry Grebe, Arjan Kuijper, Kiran Raja, Gourav Gupta, Georgios
  Zampoukis, Lazaros Tsochatzidis, Ioannis Pratikakis, S.V. Aruna~Kumar, B.S.
  Harish, Umapada Pal, Peter Peer, and Vitomir \v{S}truc.
\newblock Exploring bias in sclera segmentation models: A group evaluation
  approach.
\newblock {\em IEEE Transactions on Information Forensics and Security (TIFS)},
  18:190--205, 2023.

\bibitem{Reality_2023}
Capturing Reality.
\newblock Explore the possibilities,
  https://www.capturingreality.com/realitycapture, 2023.

\bibitem{Engine_2023}
Unreal Engine.
\newblock Point lights,
  https://docs.unrealengine.com/5.2/en-us/point-lights-in-unreal-engine/, 2023.

\bibitem{Unreal_Engine_2023}
Unreal Engine.
\newblock Post process materials,
  https://docs.unrealengine.com/5.2/en-us/post-process-materials-in-unreal-engine/,
  2023.

\bibitem{Engine_2023_utility}
Unreal Engine.
\newblock Editor utility widgets,
  https://docs.unrealengine.com/5.2/en-us/editor-utility-widgets-in-unreal-engine/,
  2023.

\bibitem{curtis2016mystery}
Alexander~E Curtis, Tanya~A Smith, Bulat~A Ziganshin, and John~A Elefteriades.
\newblock The mystery of the z-score.
\newblock {\em Aorta}, 4(04):124--130, 2016.

\bibitem{kothari2021ellseg}
Rakshit~S Kothari, Aayush~K Chaudhary, Reynold~J Bailey, Jeff~B Pelz, and
  Gabriel~J Diaz.
\newblock Ellseg: An ellipse segmentation framework for robust gaze tracking.
\newblock {\em IEEE Transactions on Visualization and Computer Graphics},
  27(5):2757--2767, 2021.

\bibitem{iandola2014densenet}
Forrest Iandola, Matt Moskewicz, Sergey Karayev, Ross Girshick, Trevor Darrell,
  and Kurt Keutzer.
\newblock Densenet: Implementing efficient convnet descriptor pyramids.
\newblock {\em arXiv preprint arXiv:1404.1869}, 2014.

\bibitem{he2016deep}
Kaiming He, Xiangyu Zhang, Shaoqing Ren, and Jian Sun.
\newblock Deep residual learning for image recognition.
\newblock In {\em Proceedings of the IEEE conference on computer vision and
  pattern recognition}, pages 770--778, 2016.

\bibitem{godard2017unsupervised}
Cl{\'e}ment Godard, Oisin Mac~Aodha, and Gabriel~J Brostow.
\newblock Unsupervised monocular depth estimation with left-right consistency.
\newblock In {\em Proceedings of the IEEE conference on computer vision and
  pattern recognition}, pages 270--279, 2017.

\bibitem{vaswani2017attention}
Ashish Vaswani, Noam Shazeer, Niki Parmar, Jakob Uszkoreit, Llion Jones,
  Aidan~N Gomez, {\L}ukasz Kaiser, and Illia Polosukhin.
\newblock Attention is all you need.
\newblock {\em Advances in neural information processing systems}, 30, 2017.

\bibitem{laina2016deeper}
Iro Laina, Christian Rupprecht, Vasileios Belagiannis, Federico Tombari, and
  Nassir Navab.
\newblock Deeper depth prediction with fully convolutional residual networks.
\newblock In {\em 2016 Fourth international conference on 3D vision (3DV)},
  pages 239--248. IEEE, 2016.

\bibitem{dierkes2018novel}
Kai Dierkes, Moritz Kassner, and Andreas Bulling.
\newblock A novel approach to single camera, glint-free 3d eye model fitting
  including corneal refraction.
\newblock In {\em Proceedings of the 2018 ACM Symposium on Eye Tracking
  Research \& Applications}, pages 1--9, 2018.

\end{thebibliography}
}

\clearpage

\onecolumn
\section*{Appendix A}

\begin{algorithm}
\caption{Eye Condition Determination for Image Extraction}
\label{alg:pre-precessing}
\begin{algorithmic}[1]

\Procedure{ImgPreProcessing}{$frames, vrAPI, model, capacity$}
    \Comment{Assume a consistent frame rate of `fps` for the input frame list}
    \State $thresholdFrames \gets \text{first } 6 \times fps \text{ frames of } frames$ 
    \Comment{Extract initial six seconds frames for threshold determination}
    
    \State $threshold \gets \Call{DetermineThreshold}{thresholdFrames, vrAPI}$
    \Comment{Compute threshold using frames from initial six seconds}

    \State $eyeImages \gets \text{empty list with capacity}$ \Comment{Initialise a list}
    \For{each $frame$ in $frames$}
        \If{\Call{isEyeOpen}{$frame, vrAPI, threshold$} and \Call{isGazeStraight}{$frame, vrAPI$}}
            \State $eyeImages.\text{append}(frame)$
            \If{$eyeImages.\text{isFull}$()}
                \State $output \gets model.\text{predict}(eyeImages)$ \Comment{Input images into the model for prediction}
                \State $eyeImages.\text{clear}$() \Comment{Clear the list for next iteration}
            \EndIf
        \EndIf
    \EndFor
    \State \textbf{return} $output$ \Comment{Return the final output from the model}
\EndProcedure

\Function{DetermineThreshold}{$initialFrames, vrAPI$}
    \State $midPointsTop \gets \text{empty list}$
    \State $midPointsBottom \gets \text{empty list}$
    \For{each $frame$ in $initialFrames$}
        \If{\Call{isGazeStraight}{$frame, vrAPI$}}
            \State $eyelidOutline \gets vrAPI.\text{getEyelidOutline}(frame)$ \Comment{Extract the eyelid outline from the current frame}
            \State $curveTop \gets \text{fitCurve}(eyelidOutline.\text{topEdge})$ \Comment{Fit a curve to the top eyelid's contour}
            \State $curveBottom \gets \text{fitCurve}(eyelidOutline.\text{bottomEdge})$ \Comment{Fit a curve to the bottom eyelid's contour}
            \State $midTop \gets \text{getMidPoint}(curveTop)$ \Comment{Calculate the midpoint of the top eyelid curve}
            \State $midBottom \gets \text{getMidPoint}(curveBottom)$ \Comment{Calculate the midpoint of the bottom eyelid curve}
            \State $midPointsTop.\text{append}(midTop)$ \Comment{Store the top eyelid midpoint for later analysis}
            \State $midPointsBottom.\text{append}(midBottom)$ \Comment{Store the bottom eyelid midpoint for later analysis}
        \EndIf
    \EndFor
    \State $threshold \gets \text{max}(midPointsTop) - \text{min}(midPointsBottom)$ \Comment{Determine the threshold}
    \State \textbf{return} $threshold$
\EndFunction

\Function{isEyeOpen}{$frame, api, threshold$}
    \State $eyelidOutline \gets api.\text{getEyelidOutline}(frame)$ \Comment{Extract eyelid outline from frame}
    \State $curveTop \gets api.\text{fitCurve}(eyelidOutline.\text{topEdge})$ \Comment{Curve fit the top eyelid}
    \State $curveBottom \gets api.\text{fitCurve}(eyelidOutline.\text{bottomEdge})$ \Comment{Curve fit the bottom eyelid}
    \State $midTop \gets api.\text{getMidPoint}(curveTop)$ \Comment{Get midpoint of the top curve}
    \State $midBottom \gets api.\text{getMidPoint}(curveBottom)$ \Comment{Get midpoint of the bottom curve}
    \If{$midBottom - midTop \ge threshold$} 
        \State \Return True \Comment{Return True if eye in frame is opened sufficiently}
    \Else 
        \State \Return False
    \EndIf
\EndFunction

\Function{isGazeStraight}{$frame, api$}
    \State $isStraight \gets api.\text{processGazeDirection}(frame)$
    \State \textbf{return} $isStraight$ \Comment{Return True if the gaze in the frame is straight, False otherwise}
\EndFunction

\end{algorithmic}
\end{algorithm}

\clearpage

\begin{figure*}[p]
    \centering
    \includegraphics[width=0.93\textwidth]{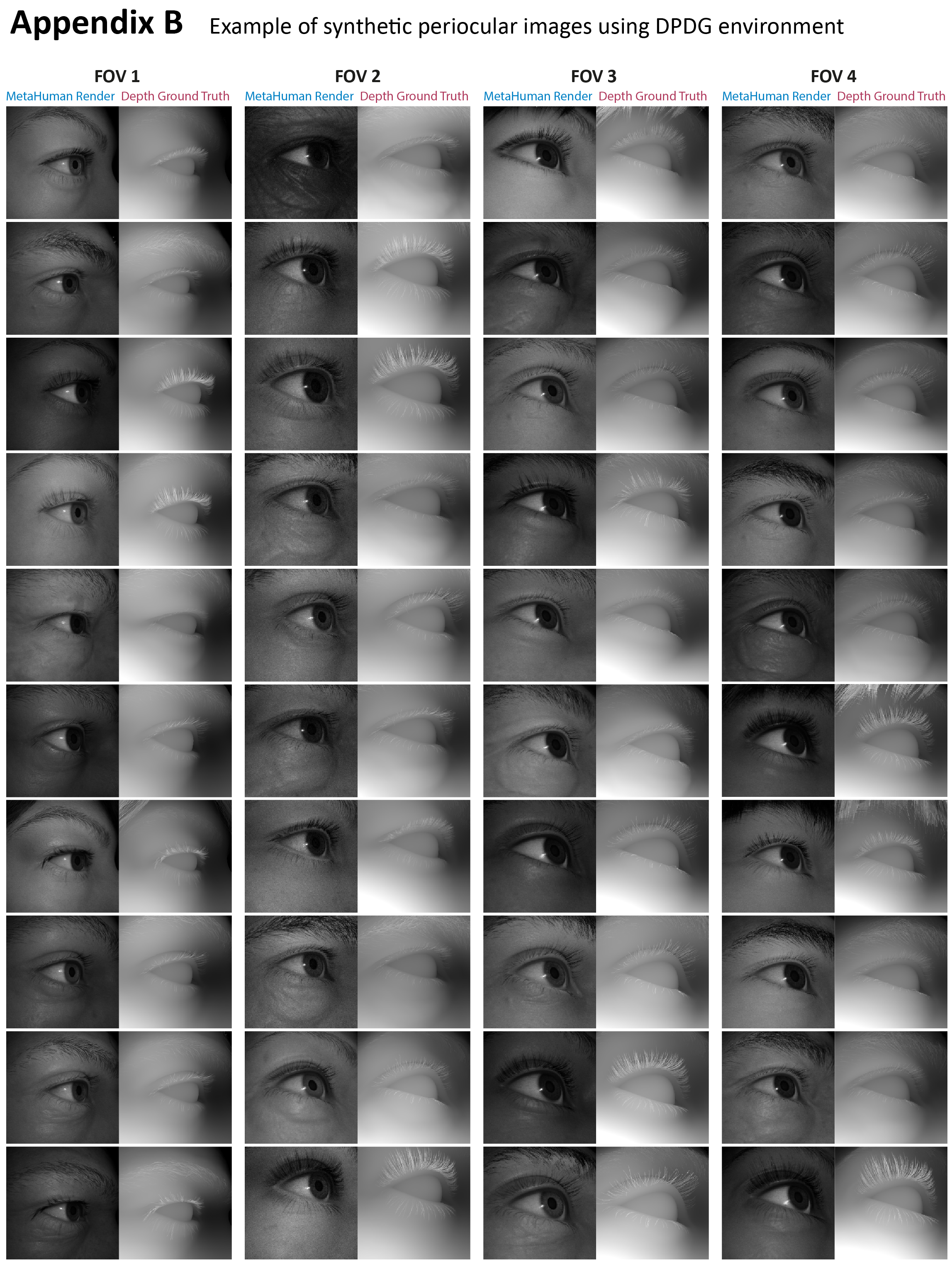}
    \label{fig:Appdix_01}
\end{figure*}

\clearpage

\begin{figure*}[p]
    \centering
    \includegraphics[width=0.84\textwidth]{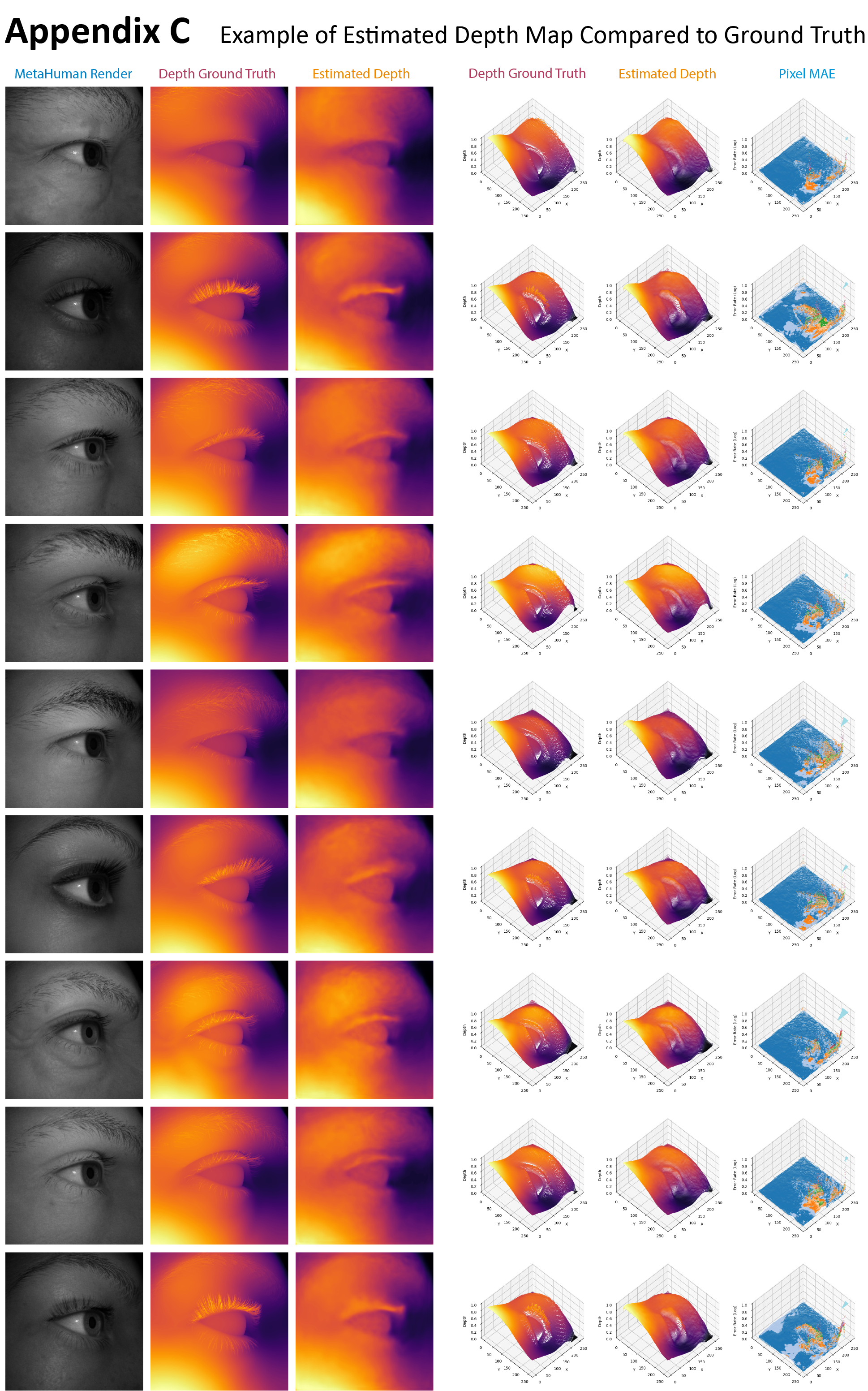}
    \label{fig:Appdix_02}
\end{figure*}

\clearpage

\begin{figure*}[p]
    \centering
    \includegraphics[width=0.82\textwidth]{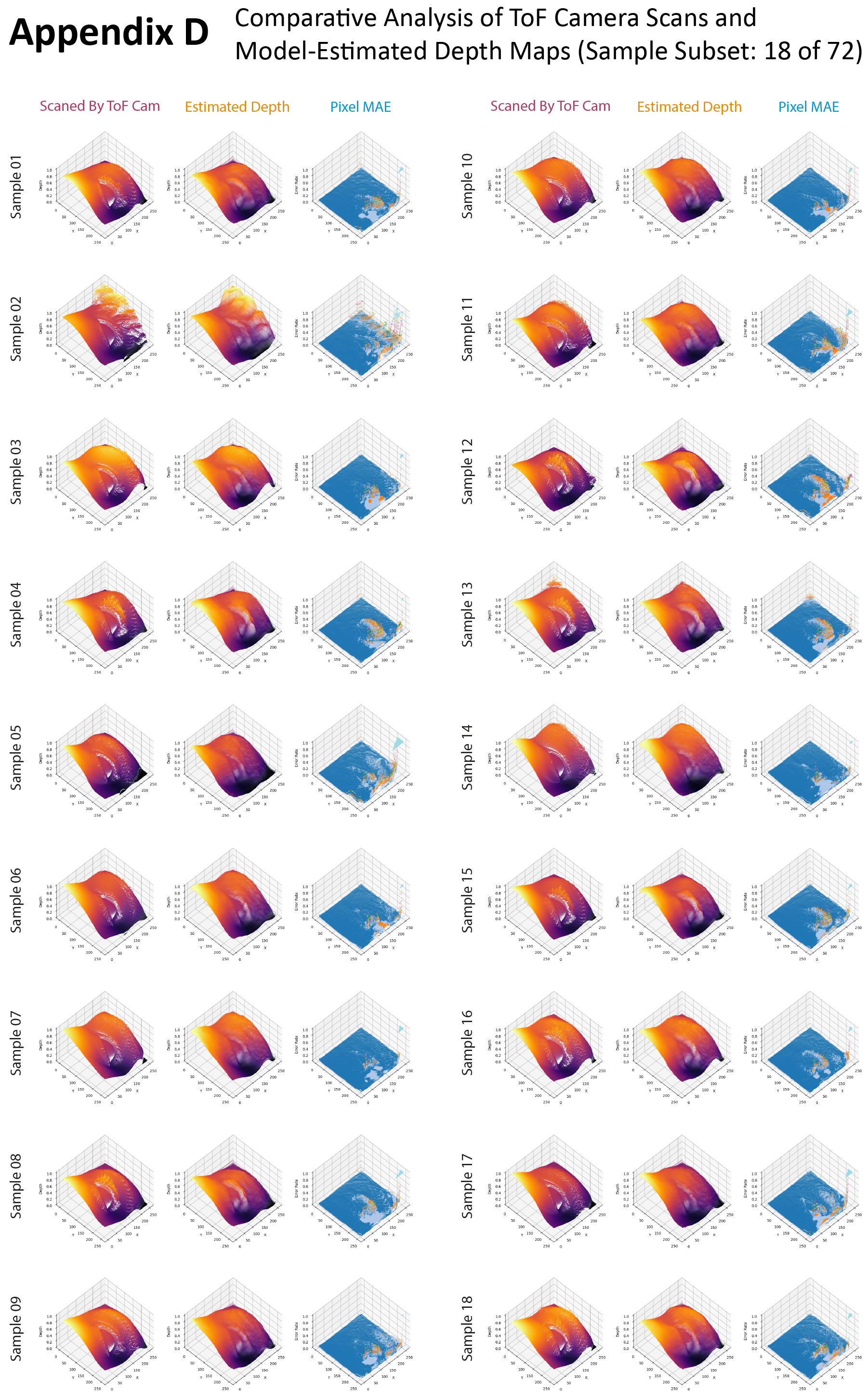}
    \label{fig:Appdix_03}
\end{figure*}

\clearpage

\twocolumn

\section*{Appendix E: Corneal Refraction Simulations using Zemax}
In our efforts to ensure the precision of pupil diameter measurements, we conducted simulations to account for the potential effects of corneal refraction. An ideal human eye model was constructed using Zemax, with the following parameters: corneal curvature \(R=8mm\), anterior chamber depth \(H=2.7mm\), refractive index of \(1.35\), and pupil diameter of \(4mm\). 

The table below presents the results from the simulation, indicating the actual pupil size, observed size when viewed externally, and the error percentage from angles ranging from \(0^\circ\) to \(60^\circ\).

\begin{table}[h]
\centering
\renewcommand{\arraystretch}{1.5}
\begin{tabular}{c c c c }
\hline
\textbf{Angle (°)} & \textbf{Actual (mm)} & \textbf{Observed (mm)} & \textbf{Error (\%)} \\
\hline
0 & 4.00 & 4.00 & 0.01 \\

10 & 4.00 & 4.01 & 0.18 \\

20 & 4.00 & 4.01 & 0.31 \\

30 & 4.00 & 4.02 & 0.52 \\

40 & 4.00 & 4.03 & 0.75 \\

50 & 4.00 & 4.04 & 0.89 \\

60 & 4.00 & 4.04 & 1.02 \\
\hline
\end{tabular}
\caption{Simulation results illustrating the effect of angle on observed pupil size.}
\label{tab:zemax_simulation}
\end{table}

From the simulation, it was determined that the maximum area differential, when observing from a frontal view to a \(60^\circ\) camera rotation, was a mere \(1.02\%\). This minute deviation justifies our assertion that refractive effects can be considered negligible for the pupil diameter measurements in our experimental setup.

\newpage
\section*{\hspace{2em}Appendix F: Glossary of Terms}
\renewcommand{\arraystretch}{1.5}
\noindent
\begin{tabularx}
{\textwidth}{@{\hspace{2em}}l X@{}}
  AI & Artificial Intelligence\\
  AbsRel & Absolute Relative Difference \\
  ANSI & American National Standards Institute \\
  API & Application Programming Interface \\
  BerHu & Reverse Huber Loss \\
  CAD & Computer-aided Design \\
  CNN & Convolutional Neural Network\\
  CIE & Commission on Illumination \\
  DES & Digital Eye Strain \\
  DPDG & Dynamic Periocular Data Generation \\
  FOV & Field of View \\
  IR LED & Infrared Light-emitting Diode \\
  RMSE & Root Mean Square Error\\
  RMSElog & Logarithmic Root Mean Square Error \\
  RGB & Red Green Blue\\
  GPU & Graphics Processing Unit\\
  MAE & Mean Absolute Error \\
  MAD & Median Absolute Deviation \\
  NDA & Non-Disclosure Agreement \\
  PPP & Preferred Practice Pattern \\
  SqRel & Squared Relative Difference \\
  ToF & Time of Flight \\
  UE & Unreal Engine \\
  VR & Virtual Reality \\
\end{tabularx}

\end{document}